%% file: main.tex
\documentclass[a4paper,fleqn]{cas-dc}
\usepackage{longtable}
\usepackage{hyperref}
\usepackage[x11names]{xcolor}
\usepackage{tabularx}
\usepackage{booktabs}
\usepackage{threeparttable}
\usepackage{array}
\usepackage{placeins}
\usepackage{soul}
\renewcommand{\hl}[1]{#1}

\usepackage[square,numbers,comma,sort&compress]{natbib}

\usepackage{caption}
\usepackage{etoolbox}
\AtBeginDocument{\captionsetup{font={rm},labelfont={rm},textfont={rm}}}
\AtBeginEnvironment{figure}{\let\sffamily\rmfamily}
\AtBeginEnvironment{table}{\let\sffamily\rmfamily}

\hypersetup{urlcolor=blue}
\let\oldprintFirstPageNotes\printFirstPageNotes
\renewcommand{\printFirstPageNotes}{%
  \begingroup
  \let\sffamily\rmfamily
  \let\footnotesize\small
  \oldprintFirstPageNotes
  \endgroup
}

\usepackage{fancyhdr}
\usepackage{float}

\let\origtwocolumn\twocolumn
\renewcommand{\twocolumn}[1][]{%
  \def\@twocolumnarg{#1}%
  \ifx\@twocolumnarg\@empty
    \origtwocolumn
  \else
    #1%
  \fi
}

\renewcommand{\textfraction}{0.2}
\renewcommand{\floatpagefraction}{0.9}

\ExplSyntaxOn
\cs_set:Npn \__cas_head:
{
  \parbox{\textwidth}{\rule{\textwidth}{0pt}}
}
\cs_set:Npn \__first_footerline:
{
  \rmfamily\small
  \ifnum\theblind>0\relax
  \else 
	\__short_authors: :~
  \fi
  { \rmfamily \itshape Preprint }
}

\cs_set:Npn \__first_foot: 
{ 
  \parbox[t]{\textwidth}
  { 
     \rule{\textwidth}{.2pt}\\
     \rmfamily\small
     \__first_footerline: \hfill \rmfamily Page~ \thepage {} ~of~ \lastpage }
}

\cs_set:Npn \__cas_foot: 
{ 
  \parbox[t]{\textwidth}
  {
   \rule{\textwidth}{.2pt}\\ 
   \rmfamily\small
   \__first_footerline:
   \hfill \rmfamily Page~\thepage {}~of~ \lastpage
  }
}
\ExplSyntaxOff

\begin{document}
\let\WriteBookmarks\relax
\def\floatpagefraction{0.9}
\def\textfraction{0.2}
\raggedbottom
\sloppy

\let\printorcid\relax

\shortauthors{Narimani et al.}
\shorttitle{~}

\title [mode = title]{Sentinel-2 for Crop Yield Estimation: A Systematic Review}

\author[1]{Mohammadreza Narimani}
\affiliation[1]{organization={University of California, Davis, Department of Biological and Agricultural Engineering},city={Davis},state={CA},country={USA}}

\author[1]{Alireza Pourreza}
\cormark[1]
\ead{apourreza@ucdavis.edu}

\author[1]{Ali Moghimi}

\author[1]{Parastoo Farajpoor}

\cortext[1]{Corresponding author}

\begin{abstract}
Accurate and timely crop yield estimation is fundamental for ensuring global food security, guiding agricultural policy, and optimizing farm management. The advent of the Copernicus Sentinel-2 satellite constellation, with its high spatial, temporal, and spectral resolutions, has catalyzed a paradigm shift in Earth observation for agriculture, enabling monitoring at the field and sub-field scale. This review synthesizes recent advancements in crop yield monitoring and estimation that leverage Sentinel-2 data. A dominant theme is the transition from regional-scale models to high-resolution, field-level assessments, driven by three primary methodological approaches: (i) empirical models using vegetation indices, increasingly coupled with machine and deep learning algorithms like Random Forest and Convolutional Neural Networks, which consistently outperform traditional regressions; (ii) the integration of process-based crop growth models (e.g., WOFOST, SAFY) through data assimilation of Sentinel-2 derived biophysical variables like Leaf Area Index (LAI); and (iii) data fusion techniques that combine Sentinel-2 optical data with weather-independent Sentinel-1 SAR imagery to overcome challenges like cloud cover. \hl{Beyond the growing use of multi-modal data, the synthesis shows that Sentinel-2-based machine learning, deep learning, and hybrid crop growth model frameworks can explain a large fraction of within-field yield variability across crops and regions, while performance is still constrained by limited and noisy ground-truth yield data, cloud-related gaps in optical imagery, and difficulties in transferring models across years and sites.} \hl{Looking ahead, the review points to tighter integration of multi-modal data streams with hybrid modeling and improved in-season ground observations as a key pathway for turning Sentinel-2-based yield estimation into robust, operational decision-support tools.} These advanced frameworks provide powerful tools for precision agriculture and sustainable intensification, offering practically relevant insights into crop performance from local to global scales.
\end{abstract}


\maketitle

\noindent
\begin{center}
\includegraphics[width=\textwidth]{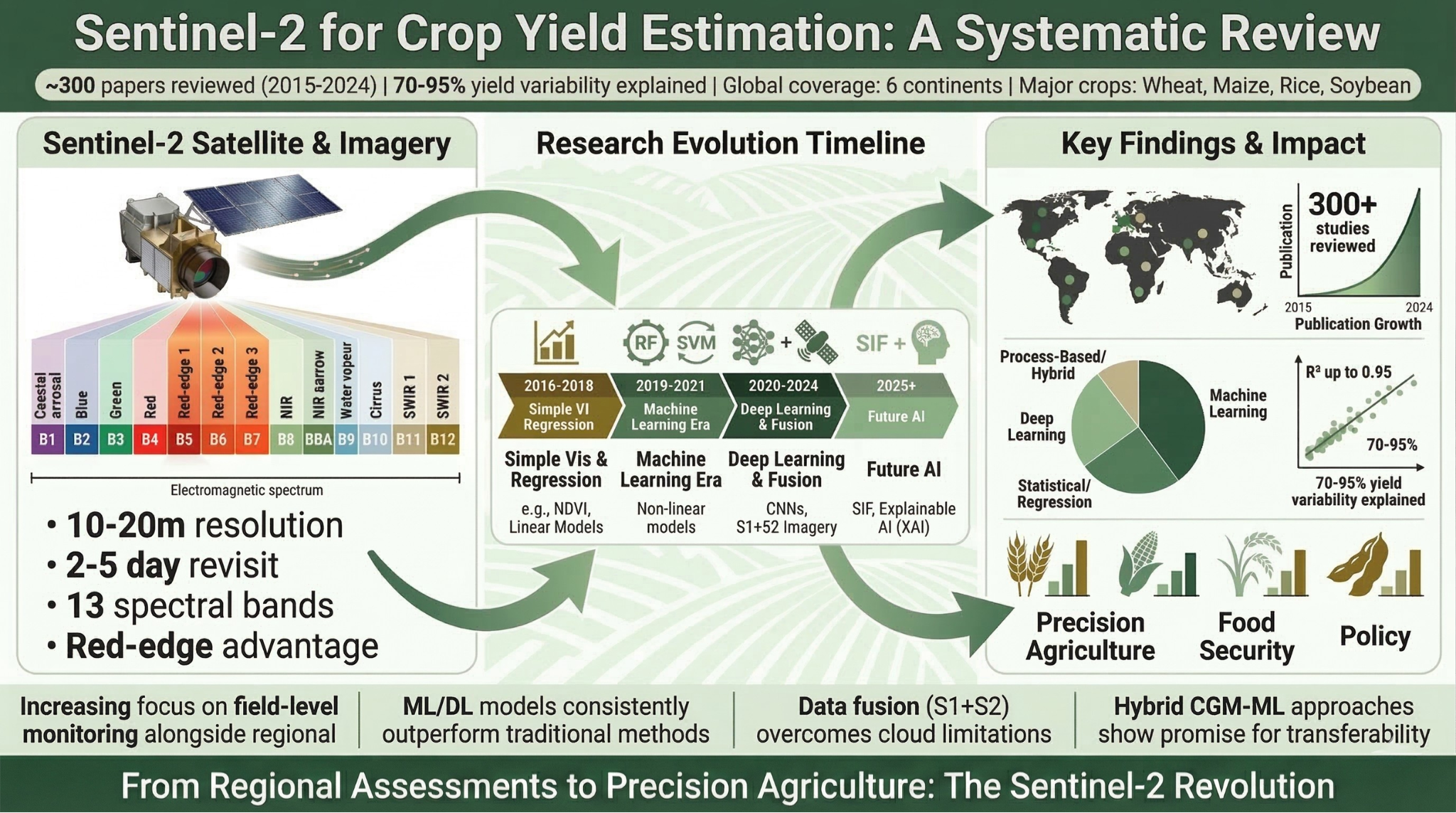}\\[5pt]
\textbf{Graphical Abstract}
\end{center}
\vspace{0.5cm}

\section{Introduction}
\label{sec:introduction}

Ensuring global food security for a growing population requires substantial increases in agricultural productivity, making timely and accurate crop yield information essential for farmers, policymakers, and global markets \cite{yu2024hidym,segarra2022farming,song2024improving}. As climate variability, input costs, and resource constraints intensify, the need for timely, spatially explicit yield information has grown, particularly to support management at the field and sub-field scale \cite{yu2024hidym,segarra2022farming,song2024improving}. Crop yield is therefore a central variable for food security planning, market stability, and on-farm decision-making. Remote sensing has long been recognized as a key technology for monitoring agricultural systems at scale and for providing consistent observations across space and time \cite{zhang2025novel,song2024improving,battude2016estimating,purnamasari2019land}. \hl{At the finest scale, the same spectral principles underpin handheld and proximal spectrometers, which enable detection of yield-related traits, plant diseases, and nutrient status at the leaf level with high precision} \cite{narimani2025ifac,farajpoor2025plant}; \hl{however, such point measurements are time-consuming to collect and difficult to generalize across entire farms. At an intermediate scale, drone-based multispectral imagery has been used for crop yield prediction, disease detection, and nutrient mapping, offering better scalability over field-sized areas} \cite{narimani2024drone,nebiker2016}. \hl{Yet at regional or national scales, drone operations are often not feasible or are limited to a small number of fields, and many applications require the broader coverage and acceptable accuracy that satellite-based monitoring can provide.} Historically, however, operational yield-related monitoring relied on coarse-resolution sensors such as MODIS and AVHRR, which, despite high temporal frequency, could not adequately resolve the small and heterogeneous fields common in many agricultural landscapes \cite{xiao2024winter,hunt2019high,qader2023exploring,zhou2020reconstruction}. While medium-resolution sensors such as Landsat improved spatial detail, long revisit intervals and frequent cloud contamination limited their ability to capture key phenological events consistently \cite{xiao2024winter,hunt2019high,qader2023exploring}.

The launch of the Copernicus Sentinel-2 satellite constellation (Sentinel-2A in 2015 and Sentinel-2B in 2017) marked a turning point for agricultural remote sensing \cite{qader2023exploring,xiao2024winter,lu2025estimation,amankulova2024integrating,narimani2025branched}. By providing freely accessible multispectral imagery with high spatial resolution (10--20 m), frequent revisit times (2--5 days at mid-latitudes), and unique spectral bands in the red-edge region, Sentinel-2 has enabled a shift from regional assessments to precise field-level and within-field yield monitoring \cite{qader2023exploring,hunt2019high,segarra2022farming}. This capability has spurred rapid methodological development and a growing body of literature. Despite this growth, the field remains fragmented across modeling paradigms, feature engineering strategies, and validation practices, making it challenging to compare findings across studies and to identify robust pathways toward operational yield estimation. No systematic review has yet synthesized this body of work with an explicit focus on crop yield estimation (rather than classification or biomass alone), leaving researchers, practitioners, and policy-makers without a consolidated reference to guide the selection of methods and future research. A systematic review that consolidates current approaches, clarifies what Sentinel-2 uniquely contributes, and highlights limitations and research gaps is therefore needed.

A major theme in the Sentinel-2 era is the advancement of empirical and data-driven models beyond simple linear regressions. Numerous studies demonstrate the utility of vegetation indices (VIs) derived from Sentinel-2, particularly those using red-edge bands such as the Normalized Difference Red-Edge Index (NDRE) and the MERIS (Medium Resolution Imaging Spectrometer) Terrestrial Chlorophyll Index (MTCI), which are less susceptible to saturation in dense canopies than the traditional Normalized Difference Vegetation Index (NDVI) \cite{jin2019smallholder,marszalek2022prediction,zhang2025novel,al2025assessment}. These VIs are increasingly used as inputs for machine learning (ML) and deep learning (DL) algorithms, including Random Forest (RF), Support Vector Machines (SVM), and neural network architectures (e.g., CNNs and LSTMs), which often outperform conventional statistical methods by capturing complex, non-linear relationships between spectral data and yield \cite{hunt2019high,mena2025adaptive,aastrom2025predicting,choudhary2022random}.

Another prominent research area is the integration of Sentinel-2 data with process-based crop growth models (CGMs) such as WOFOST, SAFY, APSIM, and CERES-Wheat \cite{tang2023estimating,qader2023exploring,zhou2020reconstruction,desloires2023out,gaso2021predicting}. Through data assimilation, satellite-derived biophysical variables---most commonly Leaf Area Index (LAI)---are used to calibrate and constrain CGM simulations, improving the accuracy of yield estimates by anchoring model dynamics to real-world observations at high spatial resolution \cite{hunt2019high,gaso2021predicting}. This hybrid approach combines mechanistic understanding of crop physiology with spatially explicit remote sensing information, enabling more robust estimations across diverse environmental conditions \cite{gaso2021predicting,tang2023estimating,song2024improving,lu2025estimation,qader2023exploring}.

Finally, data fusion strategies have emerged to overcome the inherent limitations of any single sensor. A significant trend is the synergistic use of Sentinel-2 optical data with Sentinel-1 Synthetic Aperture Radar (SAR) data \cite{jin2019smallholder,alebele2021estimation,li2025improving,gaso2024beyond}. Because SAR can penetrate clouds, this fusion supports more continuous monitoring throughout the growing season, which is particularly crucial in frequently overcast regions \cite{jin2019smallholder,yu2023improved,qader2023exploring}. Studies also show improved model performance when combining satellite imagery with ancillary data (e.g., meteorological observations, soil properties, and topographic information) to better represent the drivers of yield variability \cite{hunt2019high,mena2025adaptive,qader2023exploring,amankulova2024integrating}.

\begin{figure}[pos={!htbp}]
\centering
\includegraphics[width=\textwidth]{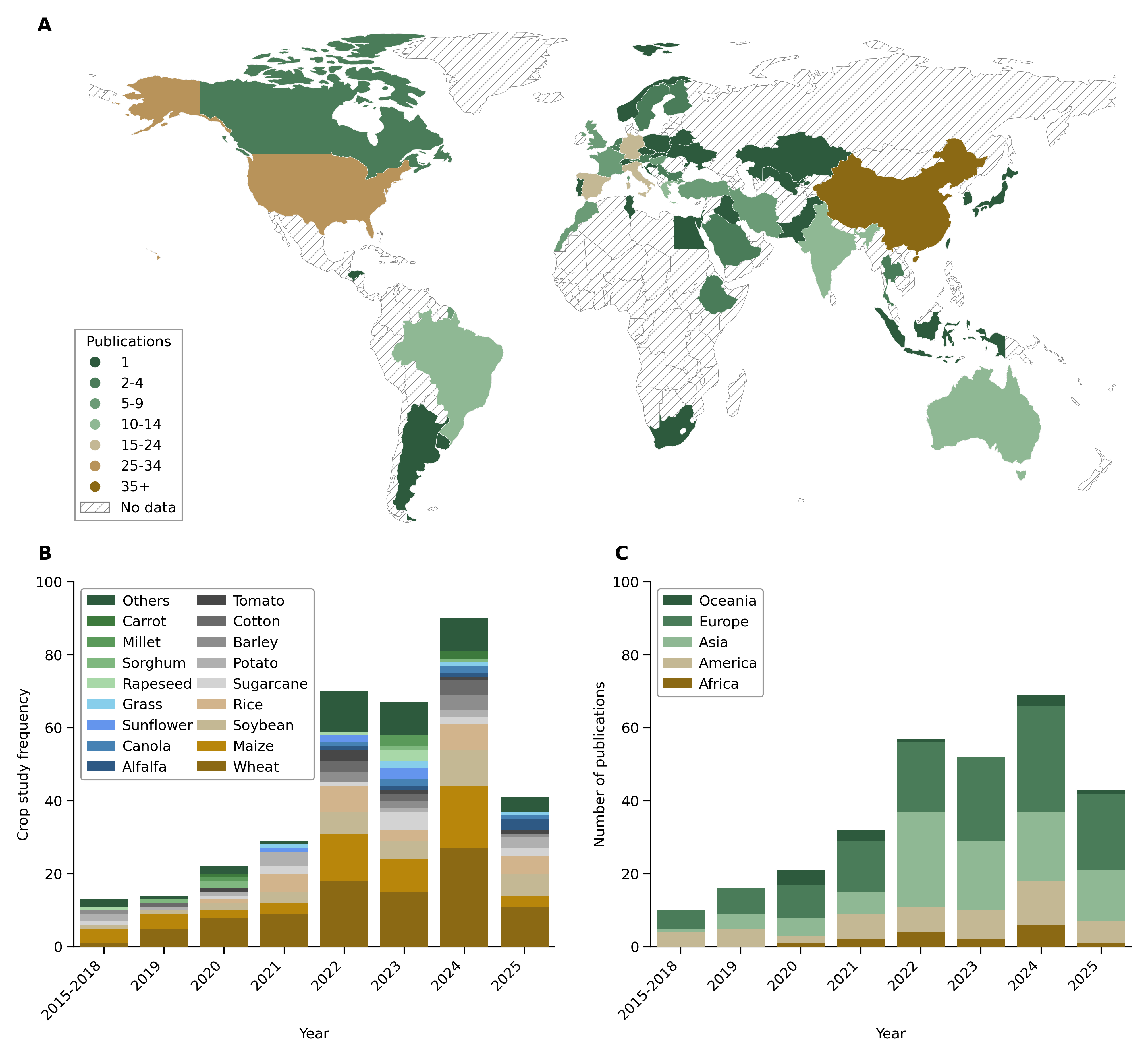}
\caption{Global trends in Sentinel-2 crop yield estimation research. (A) Geographic distribution of publications by country, showing the number of studies conducted in each nation from 2015 to present. Countries are color-coded by publication count. (B) Temporal trends in crop study frequency, displaying the evolution of research focus across major crop types over time. The stacked bars show the relative contribution of different crops to the literature each year. (C) Continental distribution of publications over time, illustrating the geographic expansion of Sentinel-2 yield estimation research across different regions. \hl{In panels B and C, the first four years (2015--2018) are merged into a single bin for consistency with the remaining year ranges.}}
\label{fig:global_trends}
\end{figure}

The objectives of this systematic review are to (i) summarize methodological trends in Sentinel-2-based crop yield estimation, (ii) compare the dominant modeling paradigms and their typical data inputs, and (iii) identify key challenges and future directions for improving accuracy, robustness, and scalability. This review synthesizes the literature around these objectives and compiles publication statistics from the papers reviewed to illustrate the rapid growth and focus of this research area (Figure~\ref{fig:global_trends}). The methods used to conduct this systematic review are described in the next section.

\vspace{0.5cm}

\section{Methods}

\subsection{Literature Search Strategy}

This systematic review synthesizes peer-reviewed literature published from 2015 to the present, coinciding with the launch of the Sentinel-2A satellite. A comprehensive search was conducted using the Web of Science Core Collection as the primary database, supplemented by searches in Scopus and Google Scholar to ensure comprehensive coverage. The search strategy employed a combination of keywords including "Sentinel-2", "crop yield", "yield estimation", "machine learning", "deep learning", "data assimilation", "wheat", "maize", "rice", "soybean", "remote sensing", and "precision agriculture". Boolean operators (AND, OR) were used to create targeted search strings that captured the intersection of Sentinel-2 satellite data and agricultural yield estimation applications.

\subsection{Inclusion and Exclusion Criteria}

Articles were included if they met the following criteria: (i) peer-reviewed research articles published in English; (ii) use of Sentinel-2 data as a primary or significant input for quantitative yield estimation of agricultural crops; (iii) focus on field-scale or regional-scale yield estimation; (iv) availability of quantitative performance metrics (e.g., R², RMSE, MAE); and (v) accessible full-text articles with valid DOIs. Studies were excluded if they: (i) focused solely on crop classification or mapping without yield estimation; (ii) examined only biomass estimation without establishing a direct link to yield; (iii) used exclusively other sensors without Sentinel-2 integration; (iv) were conference abstracts, dissertations, or non-peer-reviewed publications; or (v) were published in languages other than English.

\subsection{Study Selection and Data Extraction}

The initial search yielded 382 papers from Web of Science. A rigorous eight-step filtering process was applied to ensure quality and relevance: (1) document type screening to retain peer-reviewed research articles; (2) abstract availability verification; (3) duplicate removal based on DOI matching; (4) DOI validation and accessibility verification; (5-6) systematic accessibility screening through download attempts; (7) language filtering to retain English publications only; and (8) final quality assessment. This systematic approach resulted in 301 high-quality, accessible English research papers that form the final dataset, representing a 21.2\% reduction from the initial search that ensured comprehensive coverage while maintaining rigorous quality standards.

From each included study, the following information was systematically extracted and categorized: (i) modeling approach and algorithms used (e.g., machine learning, deep learning, statistical regression, process-based models); (ii) crop type and study region; (iii) Sentinel-2 data features utilized (e.g., spectral bands, vegetation indices, biophysical parameters); (iv) integration with other data sources (e.g., Sentinel-1 SAR, meteorological data, soil information); (v) temporal aspects (e.g., single-date vs. time-series analysis, phenological stages); (vi) reported accuracy metrics and model performance; and (vii) key findings, limitations, and recommendations.

\subsection{Data Analysis and Synthesis}

The extracted data were analyzed to identify major research themes, methodological trends, and geographic patterns. Quantitative analysis included frequency counts of modeling approaches, crop types, and geographic distribution of studies. Performance metrics were standardized where possible to enable cross-study comparisons. Temporal trends in publication patterns and methodological evolution were examined to understand the development trajectory of the field. The synthesis focused on three primary research themes: (i) empirical and machine learning approaches using vegetation indices and spectral data; (ii) integration of process-based crop growth models through data assimilation; and (iii) multi-sensor data fusion strategies combining Sentinel-2 with complementary data sources.

\subsection{Quality Assessment}

Study quality was assessed based on several criteria including: (i) clarity of methodology and reproducibility; (ii) adequacy of ground-truth data for model training and validation; (iii) appropriate use of cross-validation or independent test datasets; (iv) transparency in reporting performance metrics and limitations; and (v) consideration of spatial and temporal transferability. Studies with robust experimental designs, comprehensive validation approaches, and clear reporting of methods and results were given greater weight in the synthesis.

\vspace{0.5cm}

\section{Literature Synthesis}

Based on the search and synthesis approach described above, the following sections present the findings of the review. \hl{Abbreviations used in this review are listed in Appendix~A.}

\subsection{Remote Sensing Platforms \& Sentinel-2 Features}

The foundation of modern crop monitoring is a fleet of Earth observation (EO) satellites, each with distinct capabilities. \hl{As outlined in the introduction, earlier operational sensors either lacked the spatial detail to resolve individual fields or the temporal frequency needed to capture key phenological stages; Sentinel-2 was designed to address this gap.} The Copernicus Sentinel-2 mission, with its twin satellites (Sentinel-2A and Sentinel-2B), represents a paradigm shift for agricultural applications. Its multispectral instrument (MSI) provides data with a unique combination of technical specifications that are highly advantageous for crop monitoring \cite{desloires2023out,qader2023exploring,mahalakshmi2025soil,gamez2025alfalfa,darra2025spectral}. Key features include:

\begin{itemize}
\item \textbf{High Spatial Resolution:} \hl{Sentinel-2 offers 10 m resolution for its blue, green, red, and broadband near-infrared (NIR) bands; 20 m for its red-edge, narrow NIR, and shortwave infrared (SWIR) bands; and 60 m for atmospheric correction bands.} \cite{qader2023exploring} This 10--20 m resolution is adequate for monitoring at the field and even sub-field level, a critical improvement for precision agriculture and for studying small, fragmented farms common in many parts of the world \cite{xiao2024winter,zhou2020reconstruction,amankulova2024integrating,elders2022estimating,qader2023exploring}.

\item \textbf{High Temporal Resolution:} The two-satellite constellation (Sentinel-2A and Sentinel-2B) provides a combined revisit time of 2--5 days at mid-latitudes. \hl{A third satellite, Sentinel-2C, was launched in September 2024 and will further improve temporal resolution (shorter revisit time) as the constellation transitions to a three-satellite configuration.} This significantly increases the probability of acquiring cloud-free images during critical crop growth stages compared to platforms like Landsat \cite{xiao2024winter,zhou2020reconstruction,desloires2023out,soriano2022monitoring,segarra2022farming}. This high temporal frequency is essential for tracking rapid phenological changes, which are closely linked to final yield and thus support more accurate yield estimation \cite{qader2023exploring,dimov2022sugarcane,kolodiy2020improvement}.

\item \textbf{Unique Spectral Bands:} A key advantage of Sentinel-2 is its inclusion of multiple bands in the red-edge region (around 705 nm, 740 nm, and 783 nm) \cite{li2025improving,jin2019smallholder,prey2019simulation,mehdaoui2020exploitation,narin2022monitoring}. The red-edge is highly sensitive to vegetation chlorophyll and nitrogen content and is less prone to saturation in dense canopies compared to traditional red/NIR-based indices like NDVI \cite{prey2019simulation,mehdaoui2020exploitation,amin2025exploitation,al2025assessment,marszalek2022prediction}. This makes it particularly valuable for assessing nutrient status and estimating biomass and yield with greater accuracy \cite{jin2019smallholder,mehdaoui2020exploitation,zhang2025novel,li2025improving}.
\end{itemize}

\begin{figure}[pos={!htbp}]
\centering
\includegraphics[width=\textwidth]{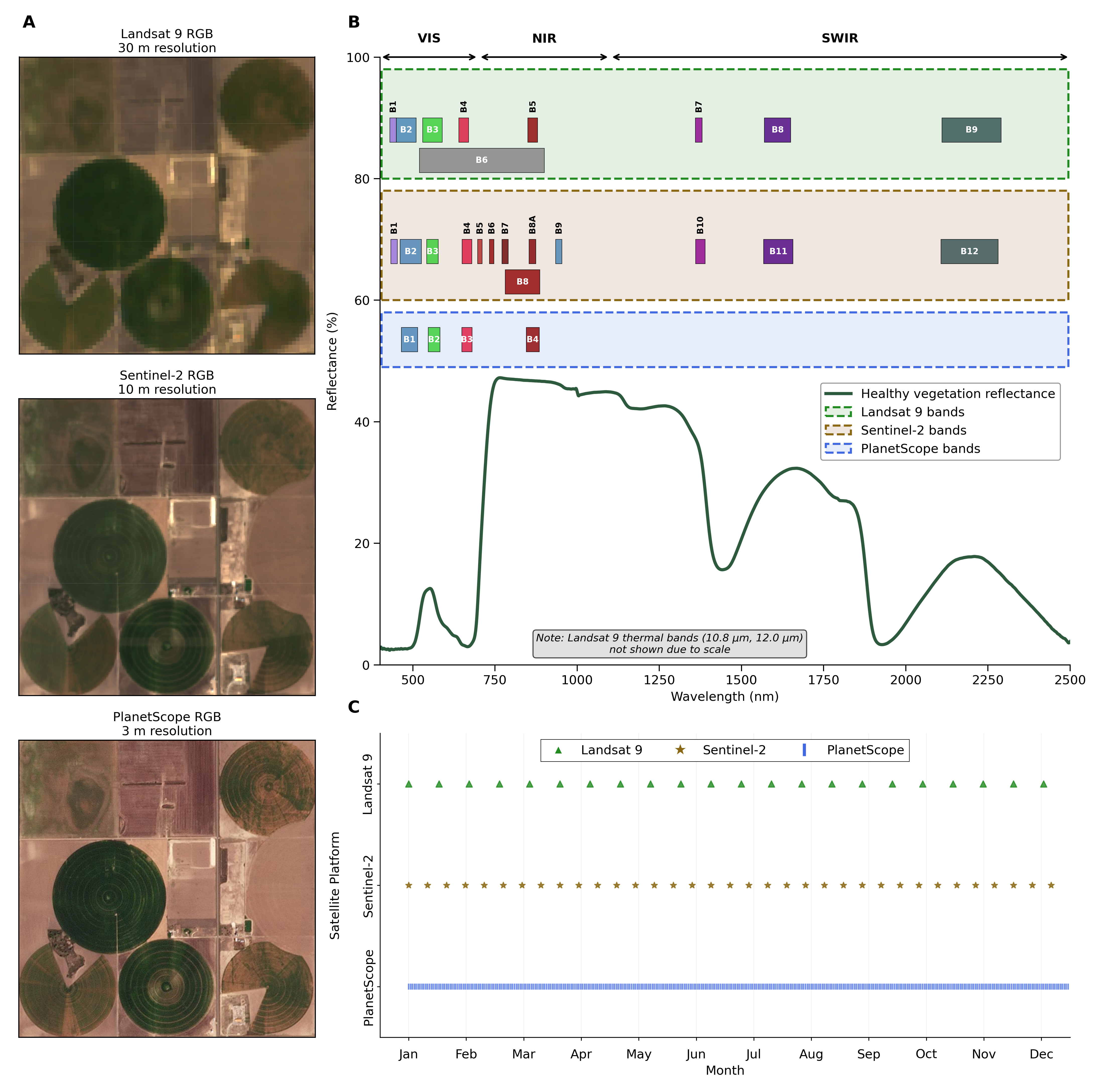}
\caption{\hl{Satellite platform comparison for crop monitoring. (A) Spatial resolution comparison showing the same agricultural area near Scott City, Kansas (38.51838°N, -100.91645°W) as captured by Landsat 9 (30 m), Sentinel-2 (10 m), and PlanetScope (3 m), demonstrating the trade-offs between spatial detail and coverage. (B) Spectral band comparison illustrating the wavelength coverage and bandwidth of each platform overlaid on a typical healthy vegetation reflectance curve (order matches panels A and C). (C) Temporal resolution comparison showing the acquisition frequency of each platform over a year, emphasizing the different revisit capabilities for continuous crop monitoring.}}
\label{fig:satellite_comparison}
\end{figure}

The availability of large-scale computational platforms for analysis, most notably Google Earth Engine (GEE), has been instrumental in enabling efficient access to and processing of the vast archives of Sentinel-2 data required for data-intensive yield estimation studies \cite{jin2019smallholder,choudhary2022random}. While Sentinel-2 offers strong capabilities as a free and open data source \cite{amankulova2024integrating,mahalakshmi2025soil,darra2025spectral,zhang2023predicting,qader2023exploring}, it is often used synergistically with other platforms. Data fusion with Sentinel-1's weather-independent Synthetic Aperture Radar (SAR) is a common strategy to overcome cloud cover limitations and ensure continuous monitoring \cite{duan2024detection,jin2019smallholder,li2025improving,gaso2024beyond,kalecinski2024crop}. Fusion with very-high-resolution commercial sensors like PlanetScope (~3 m, daily revisit) is also used to enhance spatial detail \cite{amankulova2023comparison,gogoi2023assessing,myers2021effects}, and combining Sentinel-2 with Landsat-8 data creates denser time series for more robust phenological analysis through initiatives like the Harmonized Landsat Sentinel-2 (HLS) product \cite{duan2024detection,myers2021effects,zhang2025novel,skakun2019winter}. Tables~\ref{tab:satellite_comparison} and \ref{tab:spectral_comparison} provide detailed comparisons of key satellite platforms and their spectral characteristics, while Figure~\ref{fig:satellite_comparison} illustrates the spatial, spectral, and temporal trade-offs between these systems. \hl{Agricultural monitoring advantages of Sentinel-2, Landsat 9, and PlanetScope are summarized in Appendix~B.}

\begin{table}[pos={!htbp}]
\centering
\caption{Comparison of satellite platforms for crop monitoring \cite{gatti2015sentinel,masek2020landsat,roy2021global}.}
\label{tab:satellite_comparison}
\begin{tabular*}{\textwidth}{@{\extracolsep{\fill}}llll}
\toprule
\textbf{Characteristic} & \textbf{Sentinel-2} & \textbf{Landsat 9} & \textbf{PlanetScope} \\
\midrule
Spatial Resolution & 10, 20, 60 m & 15, 30, 100 m & 3 m \\
\midrule
Temporal Resolution & 10 days (single) & 16 days & $\sim$1 day \\
& 5 days (combined) & 8 days (with L8) & \\
\midrule
Spectral Bands & 13 bands & 11 bands & 4 bands \\
\midrule
Key Spectral Features & Red-edge & Thermal IR & RGB + NIR \\
& SWIR bands & Panchromatic & \\
\midrule
Swath Width & 290 km & 185 km & 24 km \\
\midrule
Data Availability & Free & Free & Commercial \\
\midrule
Launch Date & 2015/2017 & 2021 & 2016-ongoing \\
\bottomrule
\end{tabular*}
\end{table}

\begin{table}[pos={!htbp}]
\centering
\caption{Detailed spectral band comparison for agricultural applications \cite{gatti2015sentinel,masek2020landsat,roy2021global}.}
\label{tab:spectral_comparison}
\begin{tabular*}{\textwidth}{@{\extracolsep{\fill}}llccc}
\toprule
\textbf{Spectral Region} & \textbf{Band Name} & \textbf{Sentinel-2} & \textbf{Landsat 9} & \textbf{PlanetScope} \\
\midrule
Coastal & Coastal/Aerosol & 442.7 nm (60 m) & 443 nm (30 m) & -- \\
\midrule
Visible & Blue & 492.7 nm (10 m) & 482 nm (30 m) & 490 nm (3 m) \\
& Green & 559.8 nm (10 m) & 562 nm (30 m) & 565 nm (3 m) \\
& Red & 664.6 nm (10 m) & 655 nm (30 m) & 665 nm (3 m) \\
\midrule
Red-Edge & Red-Edge 1 & 704.1 nm (20 m) & -- & -- \\
& Red-Edge 2 & 740.5 nm (20 m) & -- & -- \\
& Red-Edge 3 & 782.8 nm (20 m) & -- & -- \\
\midrule
Near-Infrared & NIR & 832.8 nm (10 m) & 865 nm (30 m) & 865 nm (3 m) \\
& Narrow NIR & 864.7 nm (20 m) & -- & -- \\
\midrule
Water Vapour & Water Vapour & 945.1 nm (60 m) & -- & -- \\
\midrule
Cirrus & Cirrus & 1373.5 nm (60 m) & 1375 nm (30 m) & -- \\
\midrule
Shortwave IR & SWIR 1 & 1613.7 nm (20 m) & 1610 nm (30 m) & -- \\
& SWIR 2 & 2202.4 nm (20 m) & 2200 nm (30 m) & -- \\
\midrule
Panchromatic & Panchromatic & -- & 590 nm (15 m) & -- \\
\midrule
Thermal IR & TIR 1 & -- & 10.8 $\mu$m (100 m) & -- \\
& TIR 2 & -- & 12.0 $\mu$m (100 m) & -- \\
\bottomrule
\end{tabular*}
\end{table}

\vspace{0.5cm}

\subsection{Modeling Approaches for Yield Estimation}

The wealth of data from Sentinel-2 has catalyzed the development and application of diverse modeling frameworks for yield estimation, which can be broadly categorized into empirical, process-based (or hybrid), and data fusion approaches.

A foundational method involves empirical statistical models, which establish direct relationships between remote sensing observations and ground-measured yield \cite{zhang2023winter,zhang2025novel}. Simple and multiple linear regressions using VIs as input features are common but are often limited by their inability to capture the complex, non-linear dynamics of crop growth \cite{corbari2022fully,zhang2023winter,choudhary2022random,segarra2022farming}.

Consequently, there has been a significant shift towards machine learning (ML) and deep learning (DL) algorithms, which now dominate the literature \cite{zhang2025novel,choudhary2022random,desloires2023out,li2025improving,dangi2025multi}. These data-driven models excel at learning complex, non-linear patterns from large datasets \cite{zhang2023winter,zhang2025novel,dangi2025multi}. Random Forest (RF) is among the most widely used algorithms, consistently demonstrating robust performance in estimating yield for various crops by integrating satellite data with environmental variables \cite{choudhary2022random,segarra2022farming,zhu2023high,khechba2025impact,jin2019smallholder}. Other popular methods include Support Vector Machines (SVM) \cite{segarra2022farming,son2022field}, Boosted Regression Trees \cite{zhu2023high,segarra2022farming}, and various neural network architectures \cite{kang2023regional,ayub2022wheat}. Advanced deep learning models, such as Convolutional Neural Networks (CNNs), Recurrent Neural Networks (RNNs) like Long Short-Term Memory (LSTM), and Graph Neural Networks (GNNs), are increasingly being used to exploit the spatio-temporal nature of satellite image time series \cite{desloires2023out,mena2025adaptive,kamangir2024large,aastrom2025predicting,dangi2025multi}. These models can process raw spectral bands directly, sometimes outperforming models based on pre-calculated VIs \cite{desloires2023out,marszalek2022prediction,perich2023pixel,cohen2025leveraging,darra2025spectral}. The implementation of these data-intensive approaches has been greatly facilitated by cloud computing platforms like GEE, which enable the processing of vast archives of Sentinel-2 data at scale \cite{marszalek2022prediction,jin2019smallholder,impollonia2024upscaling,yu2023improved,khechba2025impact}.

Hybrid approaches combine the mechanistic insights of process-based crop growth models (CGMs) with the spatial explicitness of remote sensing data through data assimilation \cite{battude2016estimating,hunt2019high,yu2024hidym,gaso2021predicting,shi2022yield}. In this framework, Sentinel-2 derived biophysical variables, most commonly LAI \cite{battude2016estimating,gaso2021predicting,shi2022yield,gaso2023efficiency,chao2021winter}, are used to calibrate or update the state variables of models like WOFOST, SAFY, APSIM, or AquaCrop \cite{battude2016estimating,hunt2019high,gaso2021predicting,shi2022yield,gaso2023efficiency}. This integration corrects the model's simulation trajectory with real-world observations, improving the accuracy and spatial detail of yield estimations \cite{hunt2019high,gaso2021predicting,gaso2023efficiency}. Some studies have also demonstrated a "hybrid-hybrid" approach, where the outputs from a data assimilation framework are fed into an ML model for a final yield estimation, which can further improve accuracy \cite{gaso2024beyond,gaso2025predicting}.

\begin{figure}[pos={!htbp}]
\centering
\includegraphics[width=\textwidth]{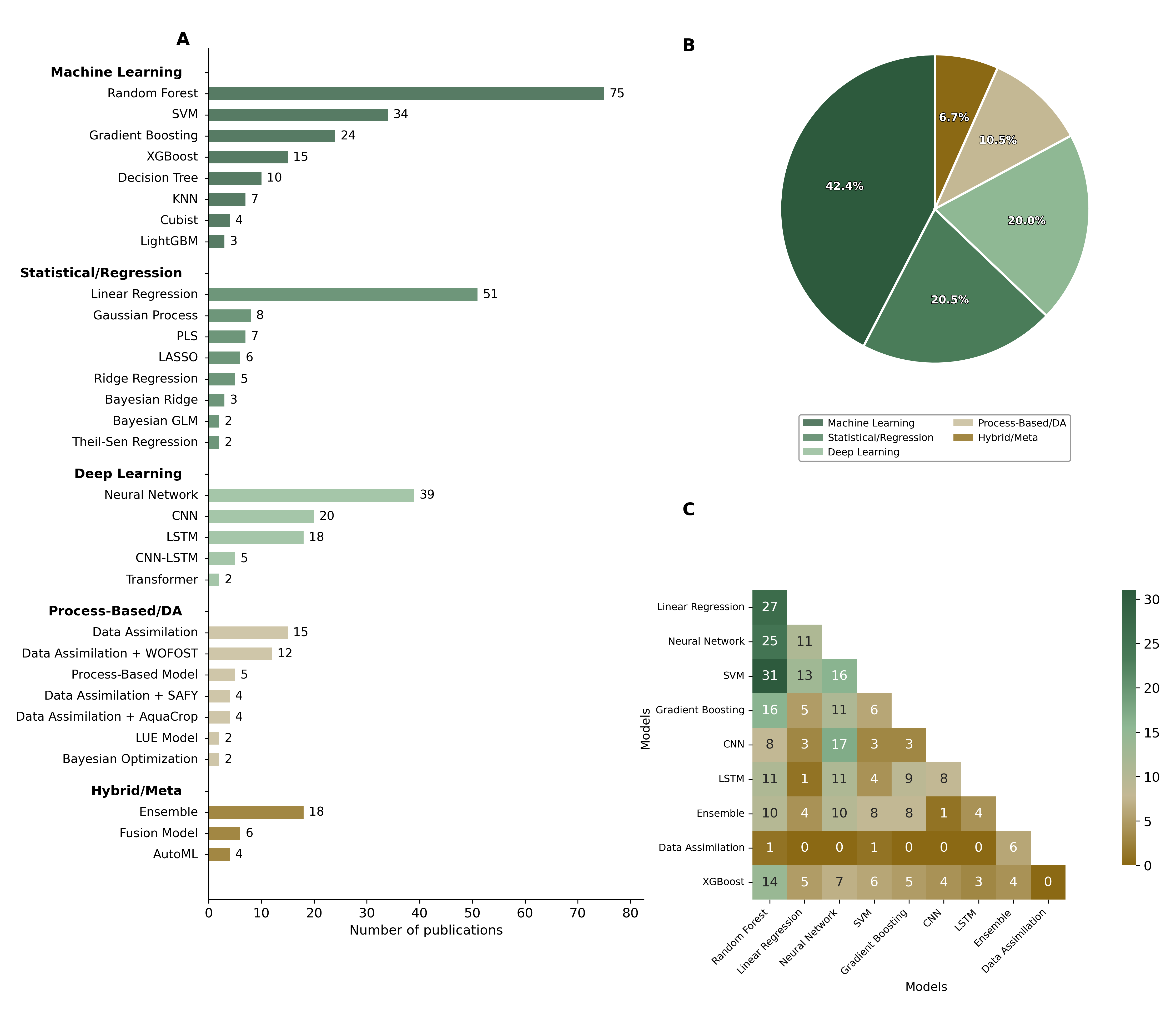}
\caption{Overview of modeling approaches for yield estimation. (A) Hierarchical bar chart showing the frequency of different modeling approaches organized by category (Machine Learning, Statistical/Regression, Deep Learning, Process-Based/Data Assimilation, and Hybrid/Meta-learning) across the reviewed studies. (B) Category distribution showing the relative prevalence of each modeling approach family in Sentinel-2 yield estimation research. (C) Co-occurrence matrix displaying the frequency of joint usage of the top 10 modeling approaches, revealing common model combinations and co-occurrence patterns in the literature.}
\label{fig:modeling_approaches}
\end{figure}

Finally, a prominent trend is the use of multi-modal data fusion within a single model \cite{mena2025adaptive,kamangir2024large}. The most common strategy involves fusing Sentinel-2 optical data with Sentinel-1 SAR data to ensure continuous observation regardless of cloud cover, which is critical in many agricultural regions \cite{duan2024detection,jin2019smallholder,li2025improving,gaso2024beyond,guillevic2024planet}. Models also frequently integrate ancillary data such as weather and climate records (e.g., temperature, precipitation), soil properties (e.g., texture, organic carbon), and topographic information (e.g., elevation, slope) to provide a more holistic representation of the drivers of yield variability \cite{choudhary2022random,khechba2025impact,mena2025adaptive,jin2019smallholder,hunt2019high}. Figure~\ref{fig:modeling_approaches} illustrates the distribution and co-occurrence patterns of these modeling approaches, while Table~\ref{tab:modeling_studies} provides a comprehensive summary of representative studies across different model categories.

\subsection{Vegetation Indices, Data Features, \& Regions}

A wide array of features derived from Sentinel-2 and other sources are used as input features in yield estimation models. The most common are Vegetation Indices (VIs), which are mathematical combinations of spectral bands designed to enhance the vegetation signal \cite{zhang2025novel,cheng2022wheat,purnamasari2019land,zhang2023winter,islam2021development}. The NDVI is the most widely used VI, often serving as a baseline for performance comparison \cite{marszalek2022prediction,segarra2022farming,ayub2022wheat,hunt2019high,schwarz2023satellite}. However, its tendency to saturate in dense canopies is a well-documented limitation \cite{marszalek2022prediction,segarra2022farming,gamez2025alfalfa,zhu2023high}.

To overcome this, researchers increasingly favor indices that leverage Sentinel-2's red-edge bands, which are more sensitive to high levels of chlorophyll and biomass \cite{prey2019simulation,mehdaoui2020exploitation,amin2025exploitation,al2025assessment,hamada2023estimating}. Prominent red-edge indices include the Normalized Difference Red-Edge Index (NDRE) \cite{khechba2025impact,desloires2023out,zhang2023season,amankulova2023sunflower,sadenova2023application}, MERIS (Medium Resolution Imaging Spectrometer) Terrestrial Chlorophyll Index (MTCI) \cite{mondschein2024mapping}, and various forms of the Chlorophyll Index Red-Edge (CIred-edge) \cite{zhang2025novel}. A novel index, the Triple Red-Edge Index (TREI), which utilizes all three of Sentinel-2's red-edge bands, has also been proposed to better capture the slope of the red-edge region \cite{al2025assessment}. Other frequently used VIs include the Green NDVI (GNDVI) \cite{amankulova2024integrating,kolodiy2020improvement,hunt2019high,segarra2022farming,schwarz2023satellite}, which can be more sensitive than NDVI in certain conditions \cite{amankulova2024integrating}, and indices that adjust for soil background effects, such as the Soil-Adjusted Vegetation Index (SAVI) \cite{amankulova2024integrating,sadenova2023application,devkota2024predicting,purnamasari2019land} and its variants. Water-sensitive indices like the Normalized Difference Water Index (NDWI) or Moisture Index (NDMI) are also used, particularly for assessing drought stress \cite{amankulova2024integrating,marszalek2022prediction,khechba2025impact,schwarz2023satellite,desloires2023out}.

Beyond VIs, models leverage a diverse set of input features. Many studies use the raw spectral reflectance bands directly as input features for ML/DL models, which can sometimes provide better performance than VIs by avoiding information loss \cite{desloires2023out,marszalek2022prediction,perich2023pixel,kamangir2024large,cohen2025leveraging}. Biophysical parameters, especially LAI, are critical features, either used as direct inputs or assimilated into CGMs \cite{battude2016estimating,gaso2021predicting,shi2022yield,gaso2023efficiency,chao2021winter}. These are often retrieved using dedicated processors like the SNAP Biophysical Processor \cite{desloires2023out,segarra2022farming,persson2024combining,zhou2020reconstruction,pelosi2022assessing}. Some studies also incorporate texture features derived from high-resolution imagery to capture canopy structure \cite{zhang2023season}. A crucial component for many models is the extraction of phenological metrics from the time series of VIs or spectral bands, such as the timing of peak greenness, growing season length, or rates of green-up and senescence \cite{dimov2022sugarcane,jin2019smallholder,duan2024detection,radovcaj2025phenology}. A more recent feature, Solar-Induced Chlorophyll Fluorescence (SIF), is gaining attention as a more direct proxy for photosynthetic activity than traditional reflectance-based VIs \cite{kang2023regional,kang2022downscaling}. \hl{SIF is retrieved from the infilling of atmospheric oxygen absorption bands in the red--near-infrared region, notably the O2-A band at $\sim$760\,nm and the O2-B band at $\sim$687\,nm, where chlorophyll fluorescence adds to the radiance measured within these absorption features.}

The reviewed literature covers a broad geographic scope, with numerous studies focusing on major agricultural regions in North America (USA, Canada) \cite{desloires2023out,mena2025adaptive,gaso2021predicting,gaso2023efficiency,zhou2020reconstruction}, Europe (e.g., Spain, France, Germany, Italy) \cite{jelinek2020landsat,qader2023exploring,segarra2022farming,impollonia2024upscaling,battude2016estimating}, Asia (e.g., China, India) \cite{xiao2024winter,tripathi2022deep,choudhary2022random,cheng2022wheat,chao2021winter}, Australia \cite{brinkhoff2024analysis,shendryk2021integrating,al2025assessment}, South America (Brazil, Argentina) \cite{mena2025adaptive,pinto2022corn,canata2021sugarcane}, and Africa (e.g., Kenya, Ethiopia, Mali) \cite{elders2022estimating,jin2019smallholder,guo2023smallholder,tiruneh2023monitoring,lobell2019sight}. The primary crops studied are cereals, particularly wheat \cite{jelinek2020landsat,yoosefzadeh2021using,qader2023exploring,segarra2022farming,cheng2022wheat} and maize (corn) \cite{desloires2023out,segarra2022farming,battude2016estimating,marszalek2022prediction,purnamasari2019land}, followed by rice \cite{choudhary2022random,impollonia2024upscaling,sayadi2025assimilating,yu2023improved,sadenova2023application}, soybean \cite{mena2025adaptive,gaso2021predicting,gaso2023efficiency,amankulova2024integrating,dado2020high}, and other important commodity and forage crops like sugarcane \cite{shendryk2021integrating,canata2021sugarcane,amaro2025performance,zhu2023high,amaro2025regional}, potato \cite{gomez2021new,gomez2019potato}, and alfalfa \cite{gamez2025alfalfa,chen2024optimal}. \hl{A comprehensive overview of the most commonly used vegetation indices in Sentinel-2 agricultural applications, including their mathematical formulations and required spectral bands, is provided in Appendix~C.}

\section{Discussion}

\subsection{Achievements, Gaps, and Challenges}

The findings synthesized above point to both achievements and challenges. The integration of Sentinel-2 data into crop monitoring has marked a significant achievement in crop yield estimation, fundamentally shifting the paradigm from coarse, regional-scale assessments to high-resolution, field- and sub-field-level estimations \cite{qader2023exploring,kayad2021ten,amankulova2024integrating,segarra2022farming}. A primary success has been the demonstrated accuracy of models built on Sentinel-2 data. Across a wide range of crops and geographies, machine learning (ML) and deep learning (DL) models have consistently achieved high coefficients of determination (R²) and low error rates, often explaining over 70--80\% of yield variability when validated against ground-truth data from combine harvesters \cite{segarra2022farming,hunt2019high,mena2025adaptive,gaso2021predicting,gaso2024beyond}. This capability enables the generation of detailed within-field yield maps, which are invaluable for precision agriculture applications like creating management zones for variable-rate fertilization and targeted irrigation \cite{kayad2021ten,tsibart2024scenarios}. Furthermore, the global coverage and free data policy of the Copernicus program have democratized access, enabling robust yield monitoring not only in large-scale commercial farming systems in Europe and North America but also in complex, heterogeneous smallholder landscapes in Africa and Asia \cite{jin2019smallholder,xiao2024winter,elders2022estimating,guo2023smallholder}.

Despite these successes, several critical gaps and challenges persist, limiting their translation into operational practice. Perhaps the most significant bottleneck is the scarcity of high-quality, high-resolution ground-truth yield data for model training and validation \cite{jin2019smallholder,kamangir2024large,elders2022estimating,guo2023smallholder}. Many studies rely on data from combine-harvester yield monitors, which are themselves prone to significant errors and require extensive, non-trivial cleaning and post-processing to be reliable \cite{elders2022estimating,segarra2022farming,perich2023pixel,girz2024within}. Alternative ground data, such as farmer surveys, can introduce bias, while precise crop cuts are labor-intensive and difficult to scale \cite{elders2022estimating,jin2019smallholder}. Such data scarcity is a major constraint for data-hungry DL models, which require large datasets to perform optimally \cite{elders2022estimating,xiao2024winter,impollonia2024upscaling,liu2025mt}. \hl{Methodological challenges specific to machine learning and deep learning models, such as overfitting when training data are limited and the limited interpretability of black-box predictions, are also widely reported and limit the operational deployment of these approaches.}

Another challenge is the inherent limitation of optical sensors. Cloud cover frequently obstructs views during critical growth periods, creating temporal gaps in Sentinel-2 time series that can compromise the accuracy of phenological metrics and yield estimations \cite{xiao2024winter,zhou2020reconstruction,zhang2025novel,soriano2022monitoring,duan2024detection}. Even with Sentinel-2's high resolution, the mixed pixel problem remains a challenge in small, fragmented fields or intercropped systems common in smallholder agriculture, where a single pixel may contain multiple crops or non-crop elements, confounding the spectral signal \cite{qader2023exploring,xiao2024winter,jin2019smallholder,elders2022estimating}.

\begin{figure}[pos={!htbp}]
\centering
\includegraphics[width=\textwidth]{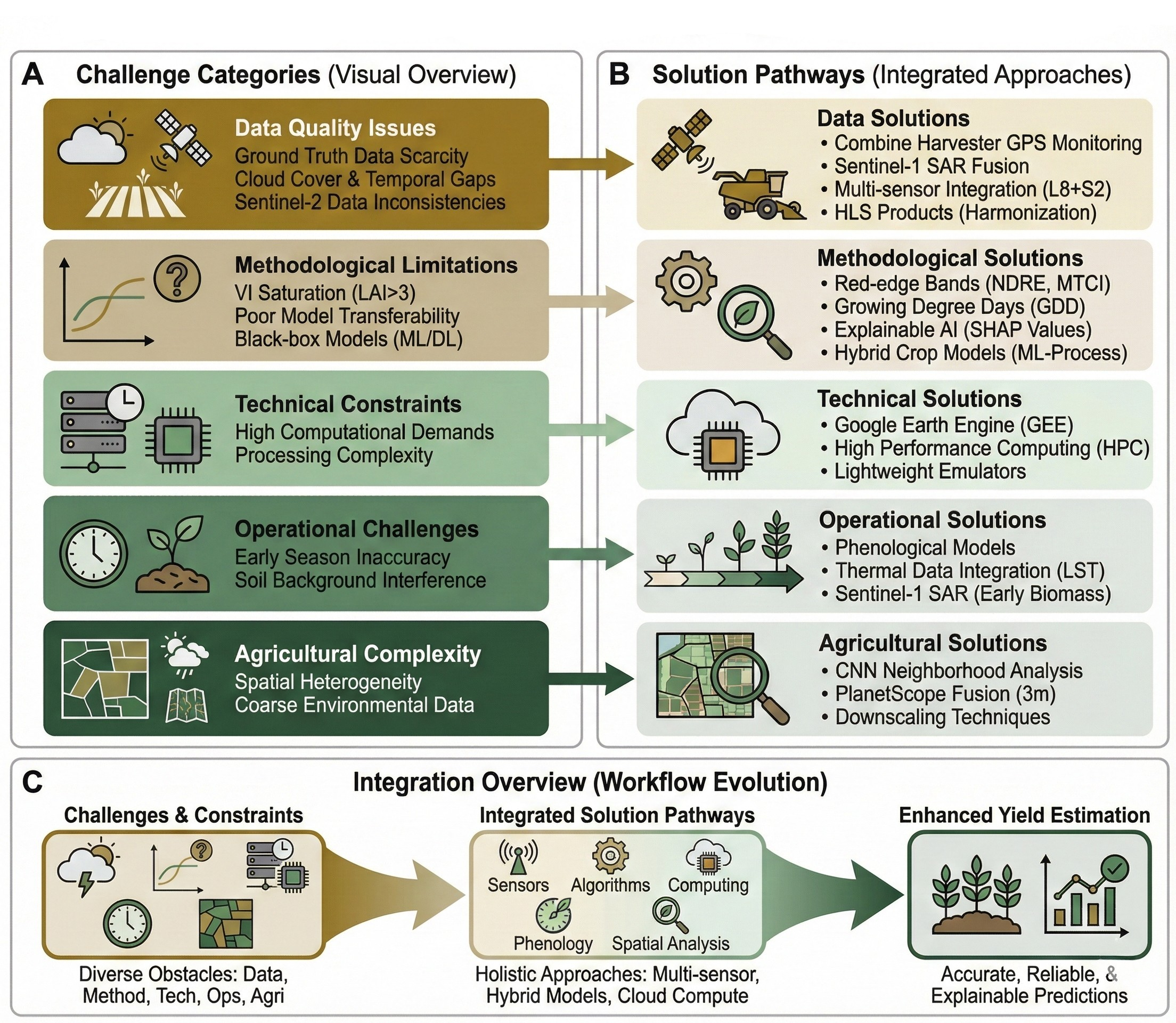}
\caption{Summary of challenges and solutions in Sentinel-2 yield estimation. A comprehensive table outlining major challenges (e.g., data availability, model transferability), their impact on model performance, and examples of solutions proposed in the literature (e.g., sensor fusion, hybrid modeling).}
\label{fig:challenges_solutions}
\end{figure}

Furthermore, the timing and transferability of yield estimations remain key hurdles. The highest accuracies are typically achieved using data from late in the growing season (e.g., reproductive or grain-filling stages), which limits the utility of estimations for in-season management decisions \cite{desloires2023out,zhang2025novel,gaso2024beyond,shendryk2021integrating,amin2024season}. Moreover, models often exhibit poor generalizability, performing well in the specific region and year for which they were trained but failing when extrapolated to new years or locations \cite{impollonia2024upscaling,desloires2023out}. This "domain mismatch" is driven by the complex interplay of genotype, environment, and management (G×E×M) interactions that vary spatially and temporally \cite{desloires2023out}, a challenge that purely data-driven models struggle to capture without diverse, multi-year training data. Figure~\ref{fig:challenges_solutions} provides a comprehensive overview of these major challenges and the corresponding solution pathways that have emerged in the literature to address them.

\vspace{0.5cm}

\FloatBarrier
\normalsize
\setlength{\LTleft}{0pt}
\setlength{\LTright}{0pt}
\setlength{\LTcapwidth}{\textwidth}
\begin{longtable}{@{\extracolsep{\fill}}p{2.9cm}p{2.7cm}p{1.9cm}p{1.7cm}p{2.1cm}p{1.7cm}@{}}
\caption{Summary of Sentinel-2 yield estimation studies}\label{tab:modeling_studies}\\[-1em]
\toprule
\textbf{Model Used} & \textbf{Data Features} & \textbf{Crop} & \textbf{Region} & \textbf{Performance} & \textbf{Reference} \\
\midrule
\endfirsthead
\multicolumn{6}{c}{{\tablename\ \thetable{} -- continued from previous page}} \\
\toprule
\textbf{Model Used} & \textbf{Data Features} & \textbf{Crop} & \textbf{Region} & \textbf{Performance} & \textbf{Reference} \\
\midrule
\endhead
\midrule \multicolumn{6}{r}{{Continued on next page}} \\
\endfoot
\bottomrule
\endlastfoot
\multicolumn{6}{l}{\textbf{\textit{Machine Learning Models}}} \\
\midrule
RF & S2 bands/VIs, Environmental data & Wheat & UK & RMSE = 0.61 t/ha & \cite{hunt2019high} \\
RF & S2 VIs (Multi-temporal, Phenological) & Sugarcane & Ethiopia & R² up to 0.84 & \cite{dimov2022sugarcane} \\
ANN, KNN, RF, SVM & S2 bands, VIs, SDD & Corn & Brazil & R² = 0.89; MAE = 0.33; RMSE = 0.42 t/ha & \cite{pinto2022corn} \\
RF, KNN, MLR, Decision Tree & S1/S2 VIs, Topographic data & Soybean & Hungary & R² = 0.41-0.89; RMSE = 0.122-0.224 t/ha & \cite{amankulova2024integrating} \\
RF, SVM, MLR & S2 VIs, LAI (RTM-derived) & Wheat & Spain & R² = 0.89; RMSE = 0.74 t/ha (RF) & \cite{segarra2022farming} \\
RF & S2 bands, VIs, SPAD values & Carrot & Saudi Arabia & R² = 0.82; RMSE = 7.8 t/ha & \cite{madugundu2024optimal} \\
RF & Multi-temporal S2/L8 (NDVI primary) & Wheat & France & R² $\geq$ 0.60; RMSE < 7 q/ha & \cite{fieuzal2020combined} \\
XGBoost & S2 VIs (EVI, NDVI), spectral bands & Legumes & Italy & R² = 0.8756 & \cite{petropoulos2025interpretable} \\
Gaussian Kernel Regression & S2 RDVI, SAR coherence & Rice & China & R² = 0.81; RMSE = 0.55 t/ha & \cite{alebele2021estimation} \\
MLR, SMR, PLS, RF, SVR & Multi-temporal S2/L8 imagery & Sugarcane & Thailand & R² = 0.79; RMSE = 3.93 t/ha (RFR) & \cite{som2024evaluating} \\
Quantile Lasso, SVM, RF & S2 red, red-edge, NIR bands & Potato & Spain & RMSE = 10.94-11.67\%; R² = 0.88-0.93 & \cite{gomez2019potato} \\
RF & S2 VIs (GNDVI) & Corn & Italy & R² = 0.48 (GNDVI); R² $\geq$ 0.6 (RF) & \cite{kayad2019monitoring} \\
Stepwise MLR, RF, GWR & S2 VIs, Topography, Rainfall & Wheat & Spain & R² = 0.83 & \cite{segarra2020estimating} \\
MLR, RF, KNN, Boosting & S2 reflectance bands, VIs (EVI, NMDI) & Durum Wheat & Greece & R² > 0.91; RMSE < 360 kg/ha (ML models) & \cite{bebie2022assessing} \\
Linear Regression, RF & S2 VIs (MTCI, GCVI, NDVI) & Maize & Ethiopia & R² up to 0.63 (RF with MTCI) & \cite{mondschein2024mapping} \\
RF & S1 SAR, S2 NDVI, SAR-NDVI & Spinach & Spain & R² = 0.89; Error = 1.4\% & \cite{mesas2025enhancing} \\
MLR, RF & S2 bands, 52 VIs, PPI & Potato & Egypt & R² = 0.734; RMSE = 1.71 kg/m² (RFR) & \cite{amin2025exploitation} \\
\midrule
\multicolumn{6}{l}{\textbf{\textit{Statistical/Regression Models}}} \\
\midrule
Bayesian MLR & S2 VIs, Climatic/Topographic data & Wheat & Iraq & R² = 0.41; RMSE = 0.698 t/ha & \cite{qader2023exploring} \\
Stepwise Linear Regression, RF & S2 time series, CWR, Climate data & Winter Wheat & Germany & R² = 0.84; RMSE = 0.56 t/ha & \cite{marszalek2022prediction} \\
Gaussian Process Regression & S2 kNDVI time series, GDD & Winter cereal & Switzerland & RMSE = 0.71 t/ha; Rel. RMSE = 7.60\% & \cite{amin2024season} \\
MLR & S2 RSIs (GNDVI) & Wheat & Morocco & R² = 0.53-0.89; RMSE = 4.29-7.78 q/ha & \cite{saad2022wheat} \\
AutoML (GLM) & S2 VIs, S1 SAR indices & Wheat & Ethiopia & R = 0.69; RMSE = 0.84-0.98 t/ha & \cite{tesfaye2022enhancing} \\
Empirical Regression & S2 NDWI composite & Winter Wheat & Sweden & MAE = 0.40 t/ha & \cite{alshihabi2024easy} \\
Empirical Regression & S2 red-edge, NIR bands & Durum wheat & Tunisia & R² = 0.55-0.73; RMSE = 3.80-4.90 q/ha & \cite{mehdaoui2020exploitation} \\
MLR & NDVI (S2/L8) & Cucumber, Bean, Corn & Honduras & Accuracy: 96.74-98.92\% & \cite{castro2024prediction} \\
Correlation-based Regression Models & L8/S2 NDVI, SAVI & Potato & Saudi Arabia & Error: 3.8-10.2\% (S2); R = 0.47-0.65 & \cite{al2016prediction} \\
Empirical Regression with Phenological Fitting & HLS VIs, Surface reflectances, AGDD & Winter Wheat & Ukraine & RMSE = 0.201 t/ha (5.4\%); R² = 0.73 & \cite{skakun2019winter} \\
Empirical Regression & S2 VIs (NDVI, MSAVI2, RCI, NDRE) & Cotton & Uzbekistan & R² up to 0.96; RMSE = 0.21 (NDRE) & \cite{khodjaev2024accuracy} \\
\midrule
\multicolumn{6}{l}{\textbf{\textit{Deep Learning Models}}} \\
\midrule
3D CNN + LSTM & S2 bands, Weather, Soil, DEM & Soybean & Argentina, Uruguay, Germany & R² = 0.86 & \cite{miranda2024multi} \\
3D CNN & S2 Surface Reflectance (B2-B4, B8) & Rice & Spain & R² = 0.92; MAE = 0.223 t/ha; MAPE = 5.78\% & \cite{fernandez2021rice} \\
3D-ResNet-BiLSTM & S2 bands, VIs, S1 SAR, Weather & Soybean & United States & R² = 0.791; RMSE = 5.56 Bu/Ac & \cite{fathi20233d} \\
\midrule
\multicolumn{6}{l}{\textbf{\textit{Process-Based/Data Assimilation Models}}} \\
\midrule
WOFOST + EnKF Data Assimilation & S1 SAR, S2 NDVI, Soil moisture & Winter Wheat & China & R² = 0.35; RMSE = 934 kg/ha (with DA) & \cite{zhuo2019assimilating} \\
SAFY + EnKF Data Assimilation & S2 time-series, LAI (from EVI2) & Winter Wheat & Sweden & R² = 0.80 (LAI); 70\% yield improvement & \cite{bouras2023wheat} \\
\midrule
\multicolumn{6}{l}{\textbf{\textit{Hybrid/Meta-Learning Models}}} \\
\midrule
Hybrid (VIs + Crop Model Stress Index) & S2 VIs, Crop Water Stress Index & Dryland Wheat & Australia & R² = 0.91; RMSE = 0.54 t/ha (combined) & \cite{zhao2020predicting} \\
\end{longtable}
\footnotesize
\textbf{Note:} Full names of abbreviations are provided in Appendix A.
\normalsize

\subsection{Outlook and Future Directions}

The future of Sentinel-2-based yield estimation is evolving towards more integrated, intelligent, and robust systems, driven by several converging trends. The continued advancement of deep learning is paramount, with a move beyond standard CNNs and LSTMs toward more sophisticated architectures. Novel approaches such as Graph Neural Networks (GNNs) are being explored to better model the irregular spatial structures of agricultural fields \cite{aastrom2025predicting}, while attention mechanisms \cite{mena2025adaptive,xiao2024winter,dangi2025multi,fathi2024mhra} and multi-task learning frameworks \cite{liu2025mt} are being used to enhance feature extraction and improve model efficiency, especially with limited ground data.

Multi-modal data fusion will become standard practice. The synergistic use of Sentinel-2 optical data with cloud-penetrating Sentinel-1 SAR imagery is already a dominant strategy for ensuring continuous, all-weather monitoring \cite{jin2019smallholder,li2025improving,yu2023improved,amankulova2024integrating,mesas2025enhancing}. The next frontier is the seamless integration of these satellite streams with ancillary data layers—including high-resolution weather forecasts, dynamic soil properties, topographic information, and management records—within unified modeling frameworks \cite{mena2025adaptive,impollonia2024upscaling,marszalek2022prediction,devkota2024predicting,gaso2025predicting}. Adaptive fusion methods, such as gated fusion networks, are emerging to dynamically weigh the contribution of each data modality, optimizing estimations based on crop type and environmental context \cite{mena2025adaptive}.

A significant paradigm shift is the move towards hybrid modeling, which couples the mechanistic understanding of process-based crop growth models (CGMs) with the pattern recognition capabilities of ML/DL \cite{gaso2024beyond,gaso2025predicting,hayman2024framework}. Instead of relying on purely empirical relationships, these hybrid systems use Sentinel-2 data to calibrate and constrain CGM simulations through data assimilation \cite{battude2016estimating,hunt2019high,jiang2025evaluation,gaso2023efficiency}. The outputs of these biophysically-grounded simulations (e.g., simulated biomass, water stress) can then serve as powerful, mechanistically-informed features for a final ML estimation model, enhancing both accuracy and interpretability \cite{gaso2024beyond}. This approach also promises to improve model transferability across different environments and years.

\begin{figure}[pos={!htbp}]
\centering
\includegraphics[width=\textwidth]{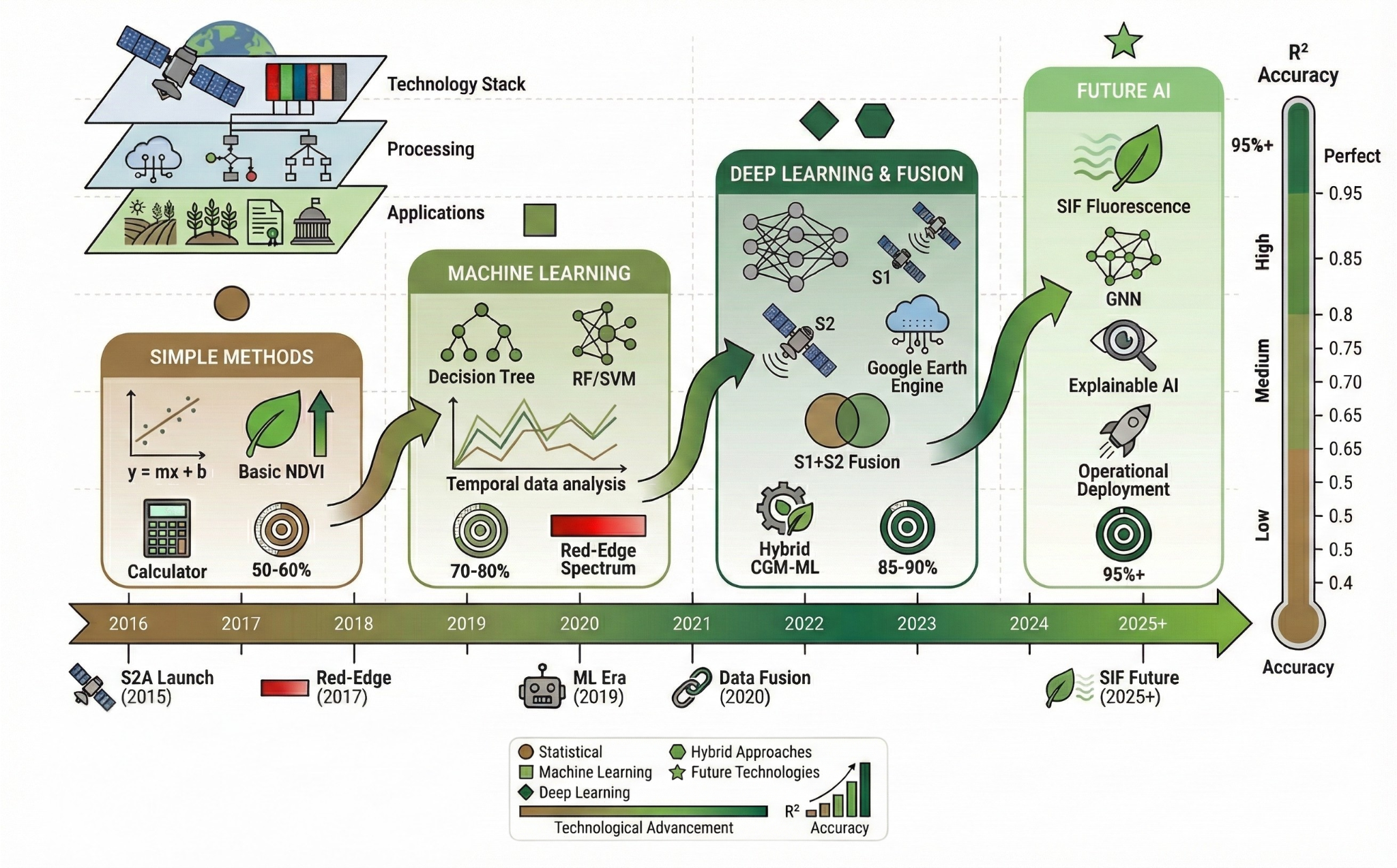}
\caption{Evolution and future outlook of Sentinel-2 yield estimation methods. A schematic timeline illustrating the progression from early methods (c. 2016) like simple VI-regressions to current advanced approaches (c. 2020--2024) using ML/DL and data fusion, and projecting future trends (c. 2025+) such as operational SIF integration and hybrid CGM-ML systems.}
\label{fig:evolution_timeline}
\end{figure}

Finally, the field will continue to benefit from the development of new data features and the sustained commitment to open data and cloud computing. There is growing interest in proxies more directly linked to plant function than traditional VIs, such as Solar-Induced Chlorophyll Fluorescence (SIF), which offers a direct measure of photosynthetic activity \cite{kang2023regional,kang2022downscaling}. As new satellite missions enhance SIF retrieval, its integration is expected to refine yield models further. All these advancements will be accelerated by platforms like GEE, which democratize access to petabyte-scale archives of satellite data and the computational power required for large-scale analytics, enabling innovation to continue at a substantial pace \cite{jin2019smallholder}. Figure~\ref{fig:evolution_timeline} illustrates this evolutionary trajectory, showing the progression from early simple regression methods to current advanced approaches and projecting future trends toward operational SIF integration and hybrid systems.

\FloatBarrier
\section{Data Availability}

The data that support the findings of this review are available in the articles cited in the reference list.

\section{Funding}

This research received no specific grant from any funding agency in the public, commercial, or not-for-profit sectors.

\section{Declaration of Interests}

The authors declare no competing interests.

\bibliographystyle{unsrt}
\bibliography{main}

\section{Appendix}

\subsection{Abbreviations}

\normalsize
\setlength{\LTleft}{0pt}
\setlength{\LTright}{0pt}
\setlength{\LTcapwidth}{\textwidth}
\begin{longtable}{@{\extracolsep{\fill}}ll@{}}
\caption{Appendix A: List of abbreviations used in this review}\label{tab:appendix_A}\\[-1em]
\toprule
\textbf{Abbreviation} & \textbf{Full Name} \\
\midrule
\endfirsthead
\multicolumn{2}{c}{{\tablename\ \thetable{} -- continued from previous page}} \\
\toprule
\textbf{Abbreviation} & \textbf{Full Name} \\
\midrule
\endhead
\midrule \multicolumn{2}{r}{{Continued on next page}} \\
\endfoot
\bottomrule
\endlastfoot
\multicolumn{2}{l}{\textbf{\textit{Machine Learning Models}}} \\
\midrule
ANN & Artificial Neural Networks \\
CNN & Convolutional Neural Network \\
GLM & Generalized Linear Model \\
KNN & K-Nearest Neighbors \\
LSTM & Long Short-Term Memory \\
MLR & Multiple Linear Regression \\
RF & Random Forest \\
RFR & Random Forest Regression \\
SMR & Stepwise Multiple Regression \\
SVM & Support Vector Machine \\
SVR & Support Vector Regression \\
\midrule
\multicolumn{2}{l}{\textbf{\textit{Statistical Methods}}} \\
\midrule
DA & Data Assimilation \\
EnKF & Ensemble Kalman Filter \\
GWR & Geographically Weighted Regression \\
PLS & Partial Least Squares \\
RTM & Radiative Transfer Model \\
\midrule
\multicolumn{2}{l}{\textbf{\textit{Crop Growth Models}}} \\
\midrule
SAFY & Simple Algorithm For Yield \\
WOFOST & World Food Studies \\
\midrule
\multicolumn{2}{l}{\textbf{\textit{Satellite Platforms}}} \\
\midrule
HLS & Harmonized Landsat Sentinel-2 \\
L8 & Landsat-8 \\
S1 & Sentinel-1 \\
S2 & Sentinel-2 \\
SAR & Synthetic Aperture Radar \\
\midrule
\multicolumn{2}{l}{\textbf{\textit{Spectral Bands}}} \\
\midrule
NIR & Near-Infrared \\
RE1 & Red-Edge 1 \\
RE2 & Red-Edge 2 \\
RE3 & Red-Edge 3 \\
SWIR1 & Shortwave Infrared 1 \\
SWIR2 & Shortwave Infrared 2 \\
\midrule
\multicolumn{2}{l}{\textbf{\textit{Biophysical Parameters}}} \\
\midrule
AGDD & Accumulated Growing Degree Days \\
CWR & Crop Water Requirements \\
DEM & Digital Elevation Model \\
GDD & Growing Degree Days \\
LAI & Leaf Area Index \\
PPI & Potato Productivity Index \\
RSIs & Remote Sensing Indices \\
SPAD & Soil Plant Analysis Development \\
\midrule
\multicolumn{2}{l}{\textbf{\textit{Performance Metrics}}} \\
\midrule
Bu/Ac & Bushels per Acre \\
MAE & Mean Absolute Error \\
MAPE & Mean Absolute Percentage Error \\
Rel. RMSE & Relative Root Mean Square Error \\
RMSE & Root Mean Square Error \\
\midrule
\multicolumn{2}{l}{\textbf{\textit{Vegetation Indices}}} \\
\midrule
EVI & Enhanced Vegetation Index \\
EVI2 & Enhanced Vegetation Index 2 \\
GCVI & Green Chlorophyll Vegetation Index \\
GNDVI & Green Normalized Difference Vegetation Index \\
kNDVI & kernel Normalized Difference Vegetation Index \\
MSAVI2 & Modified Soil-Adjusted Vegetation Index 2 \\
MTCI & MERIS (Medium Resolution Imaging Spectrometer) Terrestrial Chlorophyll Index \\
NDRE & Normalized Difference Red-Edge Index \\
NDVI & Normalized Difference Vegetation Index \\
NDWI & Normalized Difference Water Index \\
NMDI & Normalized Multi-band Drought Index \\
RCI & Red-edge Chlorophyll Index \\
RDVI & Red-edge Difference Vegetation Index \\
SAVI & Soil-Adjusted Vegetation Index \\
VIs & Vegetation Indices \\
ARVI & Atmospherically Resistant Vegetation Index \\
AWEI & Automated Water Extraction Index \\
BNDVI & Blue Normalized Difference Vegetation Index \\
CCCI & Canopy Chlorophyll Content Index \\
CIRE1 & Chlorophyll Index with Red-edge 1 \\
CIRE2 & Chlorophyll Index with Red-edge 2 \\
CIVE & Color Index of Vegetation Extraction \\
CVI & Chlorophyll Vegetation Index \\
DVI & Difference Vegetation Index \\
ExG & Excess Green \\
GARI & Green Atmospherically Resistant Index \\
GBM2 & Green Biomass Index 2 \\
GIPVI & Green Infrared Percentage Vegetation Index \\
GLI & Green Leaf Index \\
IPVI & Infrared Percentage Vegetation Index \\
IRECI & Inverted Red Edge Chlorophyll Index \\
ISR & Inverse Simple Ratio \\
LSWI & Land Surface Water Index \\
MCARI & Modified Chlorophyll Absorption in Reflectance Index \\
MGVRI & Modified Green Red Vegetation Index \\
MSI & Moisture Stress Index \\
MSR & Modified Simple Ratio \\
MTVI2 & Modified Triangular Vegetation Index 2 \\
NBR & Normalized Burn Ratio \\
NDBI & Normalized Difference Built-Up Index \\
NDI45 & Normalized Difference Index \\
NDII & Normalized Difference Infrared Index \\
NDII2 & Normalized Difference Infrared Index 2 \\
NDMI & Normalized Difference Moisture Index \\
NDRE1 & Normalized Difference Red Edge 1 \\
NDRE2 & Normalized Difference Red Edge 2 \\
NIRv & Near-Infrared Reflectance of Vegetation \\
NLI & Non-linear Index \\
NPCI & Normalized Pigment Chlorophyll Ratio Index \\
PSRI & Plant Senescence Reflectance Index \\
PVI & Perpendicular Vegetation Index \\
RE-PAP & Red Edge Physiological \& Architectural Parameter \\
REIP & Red-Edge Position Index \\
RENDVI & Red Edge Normalized Difference Vegetation Index \\
RGBVI & Red Green Blue Vegetation Index \\
RGVI & Rice Growth Vegetation Index \\
RI & Redness Index \\
SAVI2 & Soil Adjusted Vegetation Index 2 \\
SR & Simple Ratio \\
SRre1 & Simple Ratio Red-Edge 1 \\
SRre2 & Simple Ratio Red-Edge 2 \\
SRre3 & Simple Ratio Red-Edge 3 \\
TCARI & Transformed Chlorophyll Absorption in Reflectance Index \\
TGI & Triangular Greenness Index \\
TVI & Transformed Vegetation Index \\
VARI & Visible Atmospherically Resistant Index \\
VDVI & Visible Band Difference Vegetation Index \\
WDRVI & Wide Dynamic Range Vegetation Index \\
WDVI & Weighted Difference Vegetation Index \\
\end{longtable}
\vspace{-\baselineskip}

\subsection{Agricultural Monitoring Advantages}

\normalsize
\setlength{\LTleft}{0pt}
\setlength{\LTright}{0pt}
\setlength{\LTcapwidth}{\textwidth}
\begin{longtable}{@{\extracolsep{\fill}}llll@{}}
\caption{Appendix B: Agricultural monitoring advantages by satellite platform}\label{tab:appendix_B}\\[-1em]
\toprule
\textbf{Application} & \textbf{Sentinel-2} & \textbf{Landsat 9} & \textbf{PlanetScope} \\
\midrule
\endfirsthead
\multicolumn{4}{c}{{\tablename\ \thetable{} -- continued from previous page}} \\
\toprule
\textbf{Application} & \textbf{Sentinel-2} & \textbf{Landsat 9} & \textbf{PlanetScope} \\
\midrule
\endhead
\multicolumn{4}{r}{{Continued on next page}} \\
\endfoot
\bottomrule
\endlastfoot
Crop Monitoring & Red-edge bands for & Long-term data & High spatial detail \\
& chlorophyll/LAI & continuity (50+ years) & for small fields \\
\midrule
Yield Estimation & Optimal spectral & Thermal data for & Daily monitoring \\
& resolution for VIs & stress detection & capability \\
\midrule
Precision Agriculture & Field-level mapping & Regional assessments & Sub-field variability \\
& (10-20 m resolution) & & (3 m resolution) \\
\midrule
Temporal Coverage & Frequent revisits & Consistent long-term & Near-daily coverage \\
& (2-5 days) & archive & ($\sim$1 day) \\
\midrule
Cost Effectiveness & Free access & Free access & Commercial licensing \\
& & & required \\
\end{longtable}
\vspace{-\baselineskip}

\input{appendix_vegetation_table}

\section*{Acknowledgement}
The authors acknowledge the support of the University of California, Davis for providing the resources necessary to conduct this research.

\section*{Declarations}
The authors declare no conflicts of interest.

\printcredits

\end{document}

%% file: appendix_vegetation_table.tex
\subsection{Vegetation Indices}

\normalsize
\setlength{\LTleft}{0pt}
\setlength{\LTright}{0pt}
\setlength{\LTcapwidth}{\textwidth}
\renewcommand{\arraystretch}{2.5}
\newcommand{\formulasize}{\displaystyle}
\begin{longtable}{@{\extracolsep{\fill}}p{1.3cm}p{6cm}p{2cm}p{1.7cm}}
\caption{Appendix C: Vegetation indices used in Sentinel-2 agricultural applications and yield estimation studies}\label{tab:vegetation_indices}\\[-1em]
\toprule
\textbf{Index} & \textbf{Formula} & \textbf{S2 Bands} & \textbf{Reference} \\
\midrule
\endfirsthead
\multicolumn{4}{c}{{\tablename\ \thetable{} -- continued from previous page}} \\
\toprule
\textbf{Index} & \textbf{Formula} & \textbf{S2 Bands} & \textbf{Reference} \\
\midrule
\endhead
\midrule \multicolumn{4}{r}{{Continued on next page}} \\
\endfoot
\bottomrule
\endlastfoot
\multicolumn{4}{l}{\textbf{\textit{Basic Vegetation Indices}}} \\
\midrule
\multicolumn{4}{@{}l}{\footnotesize\textit{(Normalized Difference Vegetation Index)}} \\[-0.5em]
NDVI & ${\formulasize\frac{\text{NIR} - \text{RED}}{\text{NIR} + \text{RED}}}$ & B8, B4 & \cite{hunt2019high} \\[0.2em]
\multicolumn{4}{@{}l}{\footnotesize\textit{(Green Normalized Difference Vegetation Index)}} \\[-0.5em]
GNDVI & ${\formulasize\frac{\text{NIR} - \text{GREEN}}{\text{NIR} + \text{GREEN}}}$ & B8, B3 & \cite{hunt2019high} \\[0.2em]
\multicolumn{4}{@{}l}{\footnotesize\textit{(Simple Ratio)}} \\[-0.5em]
SR & ${\formulasize\frac{\text{NIR}}{\text{RED}}}$ & B8, B4 & \cite{hunt2019high} \\[0.2em]
\multicolumn{4}{@{}l}{\footnotesize\textit{(Difference Vegetation Index)}} \\[-0.5em]
DVI & ${\formulasize\text{NIR} - \text{RED}}$ & B8, B4 & \cite{hamada2021remote} \\[0.2em]
\multicolumn{4}{@{}l}{\footnotesize\textit{(Blue Normalized Difference Vegetation Index)}} \\[-0.5em]
BNDVI & ${\formulasize\frac{\text{NIR} - \text{BLUE}}{\text{NIR} + \text{BLUE}}}$ & B8, B2 & \cite{crusiol2022strategies} \\
\multicolumn{4}{@{}l}{\footnotesize\textit{(Non-linear Index)}} \\[-0.5em]
NLI & ${\formulasize\frac{\text{NIR}_{narrow} - \text{RED}}{\text{NIR}_{narrow} + \text{RED}}}$ & B8A, B4 & \cite{aslan2024artificial} \\[0.2em]
\multicolumn{4}{@{}l}{\footnotesize\textit{(Infrared Percentage Vegetation Index)}} \\[-0.5em]
IPVI & ${\formulasize\frac{\text{NIR}}{\text{NIR} + \text{RED}}}$ & B8, B4 & \cite{carneiro2023soil} \\[0.2em]
\multicolumn{4}{@{}l}{\footnotesize\textit{(Inverse Simple Ratio)}} \\[-0.5em]
ISR & ${\formulasize\frac{\text{RED}}{\text{NIR}}}$ & B4, B8 & \cite{carneiro2023soil} \\[0.2em]
\multicolumn{4}{@{}l}{\footnotesize\textit{(Green Infrared Percentage Vegetation Index)}} \\[-0.5em]
GIPVI & ${\formulasize\frac{\text{NIR}}{\text{NIR} + \text{GREEN}}}$ & B8, B3 & \cite{amin2025exploitation} \\[0.2em]
\midrule
\multicolumn{4}{l}{\textbf{\textit{Red-Edge Indices}}} \\
\midrule
\multicolumn{4}{@{}l}{\footnotesize\textit{(Normalized Difference Red-Edge Index)}} \\[-0.5em]
NDRE & ${\formulasize\frac{\text{NIR} - \text{RE1}}{\text{NIR} + \text{RE1}}}$ & B8, B5 & \cite{marszalek2022prediction} \\[0.2em]
\multicolumn{4}{@{}l}{\footnotesize\textit{(MERIS Terrestrial Chlorophyll Index)}} \\[-0.5em]
MTCI & ${\formulasize\frac{\text{RE2} - \text{RE1}}{\text{RE1} - \text{RED}}}$ & B6, B5, B4 & \cite{khechba2025impact} \\[0.2em]
\multicolumn{4}{@{}l}{\footnotesize\textit{(Chlorophyll Index with Red-edge 1)}} \\[-0.5em]
CIRE1 & ${\formulasize\frac{\text{NIR}}{\text{RE1}} - 1}$ & B8, B5 & \cite{amaro2025performance} \\[0.2em]
\multicolumn{4}{@{}l}{\footnotesize\textit{(Chlorophyll Index with Red-edge 2)}} \\[-0.5em]
CIRE2 & ${\formulasize\frac{\text{NIR}}{\text{RE2}} - 1}$ & B8, B6 & \cite{amaro2025performance} \\[0.2em]
\multicolumn{4}{@{}l}{\footnotesize\textit{(Red Edge Normalized Difference Vegetation Index)}} \\[-0.5em]
RENDVI & ${\formulasize\frac{\text{NIR} - \text{RE2}}{\text{NIR} + \text{RE2}}}$ & B8, B6 & \cite{li2025improving} \\[0.2em]
\multicolumn{4}{@{}l}{\footnotesize\textit{(Normalized Difference Index)}} \\[-0.5em]
NDI45 & ${\formulasize\frac{\text{RE1} - \text{RED}}{\text{RE1} + \text{RED}}}$ & B5, B4 & \cite{narin2022monitoring} \\[0.2em]
\multicolumn{4}{@{}l}{\footnotesize\textit{(Simple Ratio Red-Edge 1)}} \\[-0.5em]
SRre1 & ${\formulasize\frac{\text{NIR}}{\text{RE1}}}$ & B8, B5 & \cite{narin2022monitoring} \\[0.2em]
\multicolumn{4}{@{}l}{\footnotesize\textit{(Simple Ratio Red-Edge 2)}} \\[-0.5em]
SRre2 & ${\formulasize\frac{\text{NIR}}{\text{RE2}}}$ & B8, B6 & \cite{narin2022monitoring} \\[0.2em]
\multicolumn{4}{@{}l}{\footnotesize\textit{(Simple Ratio Red-Edge 3)}} \\[-0.5em]
SRre3 & ${\formulasize\frac{\text{NIR}}{\text{RE3}}}$ & B8, B7 & \cite{narin2022monitoring} \\[0.2em]
\multicolumn{4}{@{}l}{\footnotesize\textit{(Normalized Difference Red Edge 1)}} \\[-0.5em]
NDRE1 & ${\formulasize\frac{\text{RE2} - \text{RE1}}{\text{RE2} + \text{RE1}}}$ & B6, B5 & \cite{fita2025remote} \\[0.2em]
\multicolumn{4}{@{}l}{\footnotesize\textit{(Normalized Difference Red Edge 2)}} \\[-0.5em]
NDRE2 & ${\formulasize\frac{\text{RE3} - \text{RE1}}{\text{RE3} + \text{RE1}}}$ & B7, B5 & \cite{fita2025remote} \\[0.2em]
\multicolumn{4}{@{}l}{\footnotesize\textit{(Green Biomass Index 2)}} \\[-0.5em]
GBM2 & ${\formulasize\frac{\text{RE1} - \text{GREEN}}{\text{RE1} + \text{GREEN}}}$ & B5, B3 & \cite{amin2025exploitation} \\[0.2em]
\multicolumn{4}{@{}l}{\footnotesize\textit{(Canopy Chlorophyll Content Index)}} \\[-0.5em]
CCCI & ${\formulasize\frac{NDRE}{NDVI}}$ & B8, B4, B5 & \cite{suarez2024forecasting} \\[0.2em]
\midrule
\multicolumn{4}{l}{\textbf{\textit{Soil-Adjusted Indices}}} \\
\midrule
\multicolumn{4}{@{}l}{\footnotesize\textit{(Soil-Adjusted Vegetation Index)}} \\[-0.5em]
SAVI & ${\formulasize\frac{1.5(\text{NIR} - \text{RED})}{\text{NIR} + \text{RED} + 0.5}}$ & B8, B4 & \cite{li2025improving} \\[0.2em]
\multicolumn{4}{@{}l}{\footnotesize\textit{(Modified Soil-Adjusted Vegetation Index 2)}} \\[-0.5em]
MSAVI2 & ${\formulasize(\text{NIR} + 1) - \frac{1}{2}\sqrt{(2\text{NIR} + 1)^{2} - 8(\text{NIR} - \text{RED})}}$ & B8, B4 & \cite{farmonov2023combining} \\[0.2em]
\multicolumn{4}{@{}l}{\footnotesize\textit{(Soil Adjusted Vegetation Index 2)}} \\[-0.5em]
SAVI2 & ${\formulasize\frac{\text{NIR}}{\text{RED} + 0.0070}}$ & B8, B4 & \cite{amin2025exploitation} \\[0.2em]
\multicolumn{4}{@{}l}{\footnotesize\textit{(Atmospherically Resistant Vegetation Index)}} \\[-0.5em]
ARVI & ${\formulasize\frac{\text{NIR} - (\text{RED} - \text{BLUE})}{\text{NIR} + (\text{RED} - \text{BLUE})}}$ & B8, B4, B2 & \cite{hamada2021remote} \\[0.2em]
\multicolumn{4}{@{}l}{\footnotesize\textit{(Perpendicular Vegetation Index)}} \\[-0.5em]
PVI & ${\formulasize\frac{\text{NIR} - a \times \text{RED} - b}{\sqrt{a^{2}+1}}}$ & B8, B4 & \cite{darra2023can} \\[0.2em]
\multicolumn{4}{@{}l}{\footnotesize\textit{(Weighted Difference Vegetation Index)}} \\[-0.5em]
WDVI & ${\formulasize\text{NIR} - S \times \text{RED}}$ & B8, B4 & \cite{darra2023can} \\[0.2em]
\midrule
\multicolumn{4}{l}{\textbf{\textit{Water/Moisture Indices}}} \\
\midrule
\multicolumn{4}{@{}l}{\footnotesize\textit{(Normalized Difference Water Index)}} \\[-0.5em]
NDWI & ${\formulasize\frac{\text{NIR} - \text{SWIR1}}{\text{NIR} + \text{SWIR1}}}$ & B8, B11 & \cite{marszalek2022prediction} \\[0.2em]
\multicolumn{4}{@{}l}{\footnotesize\textit{(Normalized Difference Moisture Index)}} \\[-0.5em]
NDMI & ${\formulasize\frac{\text{NIR} - \text{SWIR1}}{\text{NIR} + \text{SWIR1}}}$ & B8, B11 & \cite{amaro2025performance} \\[0.2em]
\multicolumn{4}{@{}l}{\footnotesize\textit{(Land Surface Water Index)}} \\[-0.5em]
LSWI & ${\formulasize\frac{\text{NIR} - \text{SWIR1}}{\text{NIR} + \text{SWIR1}}}$ & B8, B11 & \cite{xiao2024winter} \\[0.2em]
\multicolumn{4}{@{}l}{\footnotesize\textit{(Normalized Difference Infrared Index)}} \\[-0.5em]
NDII & ${\formulasize\frac{\text{NIR} - \text{SWIR1}}{\text{NIR} + \text{SWIR1}}}$ & B8, B11 & \cite{crusiol2022strategies} \\[0.2em]
\multicolumn{4}{@{}l}{\footnotesize\textit{(Normalized Difference Infrared Index 2)}} \\[-0.5em]
NDII2 & ${\formulasize\frac{\text{NIR} - \text{SWIR2}}{\text{NIR} + \text{SWIR2}}}$ & B8, B12 & \cite{crusiol2022strategies} \\[0.2em]
\multicolumn{4}{@{}l}{\footnotesize\textit{(Normalized Burn Ratio)}} \\[-0.5em]
NBR & ${\formulasize\frac{\text{NIR} - \text{SWIR2}}{\text{NIR} + \text{SWIR2}}}$ & B8, B12 & \cite{bhumiphan2023estimation} \\[0.2em]
\multicolumn{4}{@{}l}{\footnotesize\textit{(Normalized Multi-band Drought Index)}} \\[-0.5em]
NMDI & ${\formulasize\frac{\text{NIR} - (\text{SWIR1} - \text{SWIR2})}{\text{NIR} + (\text{SWIR1} - \text{SWIR2})}}$ & B8, B11, B12 & \cite{aslan2024artificial} \\[0.2em]
\multicolumn{4}{@{}l}{\footnotesize\textit{(Moisture Stress Index)}} \\[-0.5em]
MSI & ${\formulasize\frac{\text{SWIR1}}{\text{NIR}}}$ & B11, B8 & \cite{khechba2025impact} \\[0.2em]
\multicolumn{4}{@{}l}{\footnotesize\textit{(Automated Water Extraction Index)}} \\[-0.5em]
AWEI & ${\formulasize 4(\text{GREEN} - \text{SWIR2}) - (0.25\text{NIR} + 2.75\text{SWIR2})}$ & B3, B12, B8 & \cite{aslan2024artificial} \\[0.2em]
\midrule
\multicolumn{4}{l}{\textbf{\textit{Enhanced/Modified Indices}}} \\
\midrule
\multicolumn{4}{@{}l}{\footnotesize\textit{(Enhanced Vegetation Index)}} \\[-0.5em]
EVI & ${\formulasize\frac{2.5(\text{NIR} - \text{RED})}{\text{NIR} + 6\text{RED} - 7.5\text{BLUE} + 1}}$ & B8, B4, B2 & \cite{li2025improving} \\[0.2em]
\multicolumn{4}{@{}l}{\footnotesize\textit{(Enhanced Vegetation Index 2)}} \\[-0.5em]
EVI2 & ${\formulasize\frac{2.5(\text{NIR} - \text{RED})}{\text{NIR} + 2.4\text{RED} + 1}}$ & B8, B4 & \cite{zhao2020predicting} \\[0.2em]
\multicolumn{4}{@{}l}{\footnotesize\textit{(kernel Normalized Difference Vegetation Index)}} \\[-0.5em]
kNDVI & ${\formulasize\tanh\left(\left(\frac{\text{NIR} - \text{RED}}{\text{NIR} + \text{RED}}\right)^{2}\right)}$ & B8, B4 & \cite{li2025improving} \\[0.2em]
\multicolumn{4}{@{}l}{\footnotesize\textit{(Near-Infrared Reflectance of Vegetation)}} \\[-0.5em]
NIRv & ${\formulasize\frac{\text{NIR}(\text{NIR} - \text{RED})}{\text{NIR} + \text{RED}}}$ & B8, B4 & \cite{li2025improving} \\[0.2em]
\multicolumn{4}{@{}l}{\footnotesize\textit{(Wide Dynamic Range Vegetation Index)}} \\[-0.5em]
WDRVI & ${\formulasize\frac{0.1 \times \text{NIR} - \text{RED}}{0.1 \times \text{NIR} + \text{RED}}}$ & B8, B4 & \cite{fathi2024mhra} \\[0.2em]
\multicolumn{4}{@{}l}{\footnotesize\textit{(Red-edge Difference Vegetation Index)}} \\[-0.5em]
RDVI & ${\formulasize\frac{\text{NIR} - \text{RED}}{\sqrt{\text{NIR} + \text{RED}}}}$ & B8, B4 & \cite{madugundu2024optimal} \\[0.2em]
\multicolumn{4}{@{}l}{\footnotesize\textit{(Modified Simple Ratio)}} \\[-0.5em]
MSR & ${\formulasize\frac{\frac{\text{NIR}}{\text{RED}} - 1}{\sqrt{\frac{\text{NIR}}{\text{RED}} + 1}}}$ & B8, B4 & \cite{gamez2025alfalfa} \\[0.2em]
\multicolumn{4}{@{}l}{\footnotesize\textit{(Modified Triangular Vegetation Index 2)}} \\[-0.5em]
MTVI2 & ${\formulasize\frac{1.5|1.2(\text{NIR} - \text{GREEN}) - 2.5(\text{RED} - \text{GREEN})|}{\sqrt{(2\text{NIR} + 1)^{2} - (6\text{NIR} - 5\sqrt{\text{RED}}) - 0.5}}}$ & B8, B3, B4 & \cite{amankulova2023comparison} \\[0.2em]
\midrule
\multicolumn{4}{l}{\textbf{\textit{Chlorophyll/Biomass Indices}}} \\
\midrule
\multicolumn{4}{@{}l}{\footnotesize\textit{(Green Chlorophyll Vegetation Index)}} \\[-0.5em]
GCVI & ${\formulasize\frac{\text{NIR}}{\text{GREEN}} - 1}$ & B8, B3 & \cite{hunt2019high} \\[0.2em]
\multicolumn{4}{@{}l}{\footnotesize\textit{(Chlorophyll Vegetation Index)}} \\[-0.5em]
CVI & ${\formulasize\text{NIR} \times \frac{\text{RED}}{\text{GREEN}^{2}}}$ & B8, B4, B3 & \cite{aslan2024artificial} \\[0.2em]
\multicolumn{4}{@{}l}{\footnotesize\textit{(Modified Chlorophyll Absorption in Reflectance Index)}} \\[-0.5em]
MCARI & ${\formulasize(\text{RE1} - \text{RED} - 0.2(\text{RE1} - \text{GREEN})) \times \frac{\text{RE1}}{\text{RED}}}$ & B5, B4, B3 & \cite{abreu2022estimating} \\[0.2em]
\multicolumn{4}{@{}l}{\footnotesize\textit{(Transformed Chlorophyll Absorption in Reflectance Index)}} \\[-0.5em]
TCARI & ${\formulasize 3((\text{RE1} - \text{RED}) - 0.2(\text{RE1} - \text{GREEN}) \times \frac{\text{RE1}}{\text{RED}})}$ & B5, B4, B3 & \cite{abreu2022estimating} \\[0.2em]
\midrule
\multicolumn{4}{l}{\textbf{\textit{Visible Light Indices}}} \\
\midrule
\multicolumn{4}{@{}l}{\footnotesize\textit{(Excess Green)}} \\[-0.5em]
ExG & ${\formulasize 2 \times \text{GREEN} - \text{RED} - \text{BLUE}}$ & B3, B4, B2 & \cite{pejak2022soya} \\[0.2em]
\multicolumn{4}{@{}l}{\footnotesize\textit{(Visible Atmospherically Resistant Index)}} \\[-0.5em]
VARI & ${\formulasize\frac{\text{GREEN} - \text{RED}}{\text{GREEN} + \text{RED} - \text{BLUE}}}$ & B3, B4, B2 & \cite{hamada2021remote} \\[0.2em]
\multicolumn{4}{@{}l}{\footnotesize\textit{(Triangular Greenness Index)}} \\[-0.5em]
TGI & ${\formulasize -0.5(190(\text{RED} - \text{GREEN}) - 120(\text{RED} - \text{BLUE}))}$ & B4, B3, B2 & \cite{segarra2022farming} \\[0.2em]
\multicolumn{4}{@{}l}{\footnotesize\textit{(Green Leaf Index)}} \\[-0.5em]
GLI & ${\formulasize\frac{(\text{GREEN} - \text{RED})(\text{GREEN} - \text{BLUE})}{2\text{GREEN} + \text{RED} + \text{BLUE}}}$ & B3, B4, B2 & \cite{pejak2022soya} \\[0.2em]
\multicolumn{4}{@{}l}{\footnotesize\textit{(Color Index of Vegetation Extraction)}} \\[-0.5em]
CIVE & ${\formulasize 0.441\text{RED} - 0.811\text{GREEN} + 0.385\text{BLUE} + 18.78745}$ & B4, B3, B2 & \cite{pejak2022soya} \\[0.2em]
\multicolumn{4}{@{}l}{\footnotesize\textit{(Visible Band Difference Vegetation Index)}} \\[-0.5em]
VDVI & ${\formulasize\frac{2\text{GREEN} - \text{BLUE} - \text{RED}}{2\text{GREEN} + \text{BLUE} + \text{RED}}}$ & B3, B2, B4 & \cite{nihar2023satellite} \\[0.2em]
\multicolumn{4}{@{}l}{\footnotesize\textit{(Redness Index)}} \\[-0.5em]
RI & ${\formulasize\frac{\text{RED} - \text{GREEN}}{\text{RED} + \text{GREEN}}}$ & B4, B3 & \cite{aleman2023modelling} \\[0.2em]
\multicolumn{4}{@{}l}{\footnotesize\textit{(Normalized Pigment Chlorophyll Ratio Index)}} \\[-0.5em]
NPCI & ${\formulasize\frac{\text{RED} - \text{BLUE}}{\text{RED} + \text{BLUE}}}$ & B4, B2 & \cite{amin2025exploitation} \\[0.2em]
\multicolumn{4}{@{}l}{\footnotesize\textit{(Modified Green Red Vegetation Index)}} \\[-0.5em]
MGVRI & ${\formulasize\frac{\text{GREEN}^{2} - \text{RED}^{2}}{\text{GREEN}^{2} + \text{RED}^{2}}}$ & B3, B4 & \cite{amin2025exploitation} \\[0.2em]
\multicolumn{4}{@{}l}{\footnotesize\textit{(Red Green Blue Vegetation Index)}} \\[-0.5em]
RGBVI & ${\formulasize\frac{\text{GREEN}^{2} - (\text{BLUE} \times \text{RED})}{\text{GREEN}^{2} + (\text{BLUE} \times \text{RED})}}$ & B3, B2, B4 & \cite{carneiro2023soil} \\[0.2em]
\midrule
\multicolumn{4}{l}{\textbf{\textit{Specialized Indices}}} \\
\midrule
\multicolumn{4}{@{}l}{\footnotesize\textit{(Red-Edge Position Index)}} \\[-0.5em]
REIP & ${\formulasize 700 + 40 \times \frac{\frac{\text{RED} + \text{RE3}}{2} - \text{RE1}}{\text{RE2} - \text{RE1}}}$ & B4, B7, B5, B6 & \cite{marszalek2022prediction} \\[0.2em]
\multicolumn{4}{@{}l}{\footnotesize\textit{(Inverted Red Edge Chlorophyll Index)}} \\[-0.5em]
IRECI & ${\formulasize\frac{\text{RE3} - \text{RED}}{\text{RE1}/\text{RE2}}}$ & B7, B4, B5, B6 & \cite{xiao2024winter} \\[0.2em]
\multicolumn{4}{@{}l}{\footnotesize\textit{(Red Edge Physiological \& Architectural Parameter)}} \\[-0.5em]
RE-PAP & ${\formulasize\frac{\text{RE3} - \text{RE1}}{\text{RE2}} \times \frac{\text{NIR}}{\text{RED}}}$ & B7, B5, B6, B8, B4 & \cite{ntouros2025python} \\[0.2em]
\multicolumn{4}{@{}l}{\footnotesize\textit{(Green Atmospherically Resistant Index)}} \\[-0.5em]
GARI & ${\formulasize\frac{\text{NIR} - [\text{GREEN} - 1.7(\text{BLUE} - \text{RED})]}{\text{NIR} + [\text{GREEN} - 1.7(\text{BLUE} - \text{RED})]}}$ & B8, B3, B2, B4 & \cite{hamada2021remote} \\[0.2em]
\multicolumn{4}{@{}l}{\footnotesize\textit{(Rice Growth Vegetation Index)}} \\[-0.5em]
RGVI & ${\formulasize 1 - \frac{\text{BLUE} - \text{RED}}{\text{NIR} + \text{SWIR1} + \text{SWIR2}}}$ & B2, B4, B8, B11, B12 & \cite{franch2021within} \\[0.2em]
\multicolumn{4}{@{}l}{\footnotesize\textit{(Transformed Vegetation Index)}} \\[-0.5em]
TVI & ${\formulasize 0.5(120(\text{NIR} - \text{GREEN}) - 200(\text{RED} - \text{GREEN}))}$ & B8, B3, B4 & \cite{li2022maize} \\[0.2em]
\multicolumn{4}{@{}l}{\footnotesize\textit{(Plant Senescence Reflectance Index)}} \\[-0.5em]
PSRI & ${\formulasize\frac{\text{RED} - \text{BLUE}}{\text{RE2}}}$ & B4, B2, B6 & \cite{aslan2024artificial} \\[0.2em]
\multicolumn{4}{@{}l}{\footnotesize\textit{(Normalized Difference Built-Up Index)}} \\[-0.5em]
NDBI & ${\formulasize\frac{\text{SWIR1} - \text{NIR}}{\text{SWIR1} + \text{NIR}}}$ & B11, B8 & \cite{aastrom2025predicting} \\[0.2em]
\end{longtable}
\footnotesize
\textbf{Note:} Full names of abbreviations are provided in Appendix A.
\normalsize
\renewcommand{\arraystretch}{1}
\vspace{-\baselineskip}